\magnification=\magstep1
\hoffset=0.1truecm
\voffset=0.1truecm
\vsize=23.0truecm
\hsize=16.25truecm
\parskip=0.2truecm
\def\sigbar{ {\langle \sigma \rangle}} 
\def\alphabar{ { {\bar \alpha} }} 
\def\xbar{ {\langle x \rangle}} 
\def\abar{ {\bar a} } 
\def\tz{ {\tilde \zeta} } 
\def\b{ {b}}
\def\con{ { \Lambda } }
\def\const{ { \lambda } }
\def\eff{ { {\cal E}} }
\def\newpage{\vfill\eject}

\def\pp{\parshape 2 0.0truecm 16.25truecm 2truecm 14.25truecm}
%
%
\centerline{\bf A THEORY OF THE INITIAL MASS FUNCTION}
\centerline{\bf FOR STAR FORMATION IN MOLECULAR CLOUDS} 
\bigskip 
\centerline{\bf Fred C. Adams and Marco Fatuzzo$^\dagger$} 
\bigskip 
\centerline{\it Physics Department, University of Michigan} 
\centerline{\it Ann Arbor, MI 48109, USA}
\centerline{fca@umich.edu} 
\vskip 0.15truein
\centerline{and}
\vskip 0.15truein 
\centerline{\it Institute for Theoretical Physics}
\centerline{\it University of California, Santa Barbara, CA 93106}
\vskip 0.15truein
\centerline{\it $^\dagger$current address: Wesleyan College, Macon, GA 31297}
\vskip 0.4truein
\centerline{\it submitted to The Astrophysical Journal: 23 August 1995} 
\vskip 0.22truein
\centerline{\it revised: 20 November 1995} 

\vskip 0.40truein 
\centerline{\bf Abstract} 
\medskip 

We present a class of models for the initial mass function (IMF) for
stars forming within molecular clouds.  This class of models uses the
idea that stars determine their own masses through the action of
powerful stellar outflows.  This concept allows us to calculate a
semi-empirical mass formula (SEMF), which provides the transformation
between initial conditions in molecular clouds and the final masses of
forming stars.  For a particular SEMF, a given distribution of initial
conditions predicts a corresponding IMF.  In this paper, we consider
several different descriptions for the distribution of initial
conditions in star forming molecular clouds.  We first consider the
limiting case in which only one physical variable -- the effective
sound speed -- determines the initial conditions. In this limit, we
use observed scaling laws to determine the distribution of sound speed
and the SEMF to convert this distribution into an IMF.  We next
consider the opposite limit in which many different independent
physical variables play a role in determining stellar masses. 
In this limit, the central limit theorem shows that the IMF 
approaches a log-normal form.  Realistic star forming regions 
contain an intermediate number of relevant variables; we thus 
consider intermediate cases between the two limits.   
Our results show that this picture of star formation and the IMF
naturally produces stellar mass distributions that are roughly
consistent with observations.  This paper thus provides a
calculational framework to construct theoretical models of the IMF.

\bigskip 
\noindent
{\it Subject headings:} stars: formation -- ISM: clouds -- 
galaxies: formation 

\newpage 
\medskip
\centerline{\bf 1. INTRODUCTION} 
\medskip

The initial mass function (IMF) is perhaps the most important result
of the star formation process.  A detailed knowledge of the initial
mass function is required to understand galaxy formation, the chemical
evolution of galaxies, and the structure of the interstellar medium.
Unfortunately, however, the current theory of star formation says very
little about the IMF (see, e.g., the reviews of Shu, Adams, \& Lizano
1987, hereafter SAL; Zinnecker, McCaughrean, \& Wilking 1993). In
particular, we remain unable to calculate the initial mass function
from first principles.

Given the extreme importance of the IMF and the many successes of the
current theory of star formation, we feel that it is now time to begin
building models of the IMF.  The purpose of this present paper is to
present a class of IMF models which use the idea that stars, in part,
determine their own masses through the action of powerful stellar
winds and outflows (see, e.g., SAL; Lada \& Shu 1990). 
Within the context of the current theory of star formation described
below (\S 1.2), we can conceptually divide the process which
determines the IMF into two subprocesses:

\item{\bf[1]} The spectrum of initial conditions produced by 
molecular clouds (the star forming environment).  

\item{\bf[2]} The transformation between a given set of 
initial conditions and the properties of the final (formed) star.
This transformation is accomplished through the action of stellar 
winds and outflows. 

\noindent 
Notice that molecular clouds are not observed to be collapsing as a
whole; on average, the lifetime of a molecular cloud is (at least) 
an order of magnitude longer than the free-fall time (e.g., Zuckerman
\& Palmer 1974).  Thus, these clouds exhibit quasi-static behavior and
it makes sense to conceptually divide the process of determining the 
distribution of stellar masses into the two steps given above
(see Zinnecker 1989, 1990). 

A large body of previous work on the IMF exists in the literature
(see, e.g., the reviews of Zinnecker, McCaughrean, \& Wilking 1993;
Elmegreen 1985).  Many of these studies use the idea that
fragmentation of clouds leads directly to the masses of the forming
stars (e.g., Hoyle 1953; Larson 1973; Bodenheimer 1978; Elmegreen \&
Mathieu 1983). More recent work (Larson 1992, 1995) has extended these
ideas to include the observed fractal and hierarchical structure of
molecular clouds (e.g., Scalo 1985; Dickman, Horvath, \& Margulis
1990; Scalo 1990; Lada, Bally, \& Stark 1991; Houlahan \& Scalo
1992). Zinnecker (1984, 1985, 1989, 1990) has discussed the two
subprocesses given above and has explored several different
fragmentation schemes to produce the IMF.  The concept that stars help
determine their own masses has just now begun to be incorporated into
models of the IMF.  Silk (1995) has discussed the IMF for stars which
have masses limited by feedback due to both ionization and
protostellar outflows.  Nakano, Hasegawa, \& Norman (1995) have
introduced a model in which stellar masses are sometimes limited by
the mass scales of the formative medium and are sometimes limited by
feedback.  Finally, a more primitive version of this current theory
has been presented previously (Adams 1995).

For the point of view of the IMF adopted in this paper, 
traditional arguments based on the Jeans mass are not applicable.  
A characteristic feature of molecular clouds is that they are highly
non-uniform; clumpiness and structure exist on all resolvable spatial
scales.  In particular, no characteristic density exists for these
clouds and hence no (single) Jeans mass exists.  We stress that, at
least in the context of present day star formation in molecular
clouds, {\it the Jeans mass has virtually nothing to do with the
masses of forming stars}.

\bigskip
\centerline{\it 1.1 The IMF Observed} 
\medskip 

We begin this discussion by emphasizing that stars can only exist in a
finite range of masses.  Stellar objects with masses less than about
0.08 $M_\odot$ cannot produce central temperatures hot enough for the
fusion of hydrogen to take place; objects with masses less than this
hydrogen burning limit are brown dwarfs (see, e.g., Burrows, Hubbard, 
\& Lunine 1989; Burrows et al. 1993; Laughlin \& Bodenheimer 1993). 
On the other end of the possible mass range, stars with masses greater 
than about 100 $M_\odot$ cannot exist because they are unstable (e.g., 
Phillips 1994).  Thus, stars are confined to the mass range 
$$0.08 \le m  \le 100  \, , \eqno(1.1)$$
where we have defined $m \equiv M_\ast / (1 M_\odot)$. Notice that
this mass range is rather narrow in the sense that it is much smaller 
than the conceivable range of masses.  Stars form within galaxies 
which have masses of about $10^{11} M_\odot$ and stars are made up 
of hydrogen atoms which have masses of about $10^{-24}$ g $\sim$ 
$10^{-57} M_\odot$.  Thus, galaxies {\it could} build objects anywhere
in the mass range from $10^{-57} M_\odot$ to $10^{11} M_\odot$, a
factor of $10^{68}$ in mass scale. And yet, as we have discussed
above, stars live in the above mass range which allows stellar masses
to vary by only a factor of $\sim 10^3$.

The initial mass function in our galaxy has been estimated 
empirically. The first such determination (Salpeter 1955) 
showed that the number of stars with masses in the range 
$m$ to $m + dm$ is given by the power-law relation 
$$f(m) \, dm \, \sim m^{-\b} \, dm \, , \eqno(1.2)$$
where the index $\b$ = 2.35 for stars in the mass range 
$0.4 \le m \le 10$.  However, more recent work (e.g., 
Miller \& Scalo 1979; Scalo 1986; Rana 1991; Tinney 1995) 
suggests that the mass distribution deviates from a pure 
power-law. The distribution becomes flatter (and may 
even turn over) at the lowest stellar masses ($\b$ approaches 
unity for the lowest masses 0.1 $\le m \le$ 0.5) and becomes 
steeper at the highest stellar masses ($\b \sim$ 3.3 for $m >$ 10).
The observed IMF can be approximated with an analytic fit using 
a log-normal form (Miller \& Scalo 1979), i.e., 
$$\log_{10} f (\log_{10} m) = a_0 - a_1 \log_{10} m - 
a_2 (\log_{10} m)^2 \, , \eqno(1.3)$$
where $a_0$ = 1.53, $a_1$ = 0.96, and $a_2$ = 0.47. 
The true IMF has more structure than a simple log-normal 
form (Scalo 1986; Rana 1991), although equation [1.3] provides 
a good analytic reference distribution.  Figure 1 shows three 
successive approximations to the observed IMF: the Salpeter 
power-law [1.2], the Miller/Scalo log-normal form [1.3], and 
the more recent distribution taken from Table 2 of Rana (1991). 
We note that the construction of the IMF from observational 
quantities (e.g., the observed luminosity function) requires 
considerable processing. However, the basic features of the IMF
seem to be very robust.  As a general rule, the IMF does not 
change very much from one star forming region to another.  
For the sake of definiteness, in this paper, we use the 
analytic fit given by equation [1.3] as a benchmark with 
which to compare our theoretical models.

\bigskip
\centerline{\it 1.2 The Current Theory of Star Formation} 
\medskip 

In the last decade, a generally successful working paradigm of star
formation has emerged (see, e.g., SAL for a review).  Since the IMF 
models of this paper use this paradigm as a starting point, in this
section we quickly review its basic features.  One result of this
present work is thus a consistency check -- we show that this star
formation paradigm {\it can} produce an IMF similar to that observed.

In our galaxy today, star formation takes place in molecular clouds.  
These clouds thus provide the initial conditions for the star forming 
process. Molecular clouds have very complicated substructure. 
In addition, these clouds exhibit molecular linewidths $\Delta v$ 
which contain a substantial non-thermal component (e.g., Myers \& 
Fuller 1992); this linewidth broadening is generally interpreted 
as a ``turbulent'' contribution to the velocity field.  

Molecular clouds are supported against their self-gravity by both
``turbulent'' motions and by magnetic fields.  The fields gradually
diffuse outward (relative to the mass) and small centrally condensed
structures known as molecular cloud cores are formed.  These cores
represent the initial conditions for protostellar collapse. In the
simplest picture, these cores can be (roughly) characterized by two
physical variables: the effective sound speed $a_{\rm eff}$ and the 
rotation rate $\Omega$.  The effective sound speed generally
contains contributions from both magnetic fields and ``turbulence'',
as well as the usual thermal contribution.  The total effective 
sound speed can thus be written 
$$a^2_{\rm eff} = a^2_{\rm therm} + a^2_{\rm mag} + a^2_{\rm turb} \, . 
\eqno(1.4)$$

The molecular cloud cores eventually undergo dynamic collapse, which
proceeds from inside-out; in other words, the central parts of the
core fall in first and successive outer layers follow as pressure
support is lost from below (Shu 1977).  Since the infalling material
contains angular momentum (the initial state is rotating), not all of
the infalling material reaches the stellar surface.  The material with
higher specific angular momentum collects in a circumstellar disk.
The collapse flow is characterized by a well defined mass infall rate
$\dot M$, the rate at which the central object (the forming star/disk
system) gains mass from the infalling core.  Notice that no mass scale
appears in the problem, only a mass infall {\it rate}. In particular, 
the total amount of mass available to a forming star is generally 
much larger than the final mass of the star. 

One important characteristic of the rotating infalling flow 
described above is that the ram pressure of the infall is weakest 
at the rotational poles of the object.  The central star/disk system
gains mass until it is able to generate a powerful stellar wind which
breaks through the infall at the rotational poles and thereby leads
to a bipolar outflow configuration.  Although the mechanism which 
generates these winds remains under study (see the review of 
K{\" o}nigl \& Ruden 1993), the characteristics of outflow
sources have been well studied observationally (see the review of 
Lada 1985). One of the basic working hypotheses of star formation
theory is that these outflows help separate nearly formed stars from
the infalling envelope and thereby determine, in part, the final
masses of the stars (SAL; Lada \& Shu 1990).  In this paper, we use
this idea as the basis for calculating a transformation between the
initial conditions in a molecular clouds core and the final mass of
the star produced by its collapse (see \S 2; Shu, Lizano, \& Adams
1987, hereafter SLA; Adams 1995).

\bigskip
\centerline{\it 1.3 Organization of the Paper} 
\medskip 

This paper is organized as follows.  In \S 2, we derive a
semi-empirical mass formula which provides a transformation between
the initial conditions in a star forming region and the final masses
of the stars formed.  In subsequent sections, we use this
transformation in conjunction with the observed properties of
molecular clouds to derive an initial mass function. In \S 3, we first
consider the limit in which only one physical variable (the effective
sound speed) determines stellar masses.  In \S 4, we consider the
opposite limit in which a large number $n$ of physical variables
contribute to the determination of stellar masses; in the limit 
$n \to \infty$, the central limit theorem implies that the IMF
approaches a log-normal form.  In \S 5, we explore more complicated
models of the IMF; these models are intermediate between the limiting
cases studied in the two previous sections.  In this section we also
consider the effects of binary companions on the IMF. Finally, we
conclude in \S 6 with a summary and a discussion of our results.

\bigskip 
\centerline{\bf 2. A SEMI-EMPIRICAL MASS FORMULA} 
\medskip 

In this section, we calculate the transformation between initial
conditions and the final masses of the stars produced.  In other
words, we derive a semi-empirical mass formula (SEMF) for the masses
of forming stars.  In the current picture of star formation, the final
masses of stars are produced in part through the action of powerful
outflows.  Thus, we must know how this stellar outflow stops the
inflow and separates the star/disk system from its molecular
environment.  Although this process has not been well studied, we
obtain a working estimate by balancing the ram pressure of the outflow
against that of the infall. For this calculation, we adopt the
arguments first presented by SLA (see also Adams 1995).

The key concept in this argument is that the final mass of a star is
determined by the condition that the stellar outflow is strong enough
to reverse the direct infall onto the star.  We write this condition
in the form 
$$\dot M_w = \delta \dot M_\ast ,   \eqno(2.1)$$
where $\dot M_w$ is the mass loss rate of the wind and 
$\dot M_\ast$ mass infall rate onto the star itself; this 
infall rate is generally only a fraction of the total 
mass infall rate because much of the infalling material 
falls directly onto the disk. 
Notice that we should really compare the ram pressure 
$(\sim {\dot M}_w v_w)$ of the wind with that of the infall 
$(\sim {\dot M}_\ast v_\ast)$.  However, both velocities are 
determined by the depth of the stellar potential well and 
are thus comparable in magnitude; we thus divide out the 
velocities and incorporate any uncertainties into the 
parameter $\delta$.   

Since we do not yet fully understand how high velocity outflows 
are produced, we must proceed in a semi-empirical manner (although 
considerable progress in this area has recently been made -- see 
Shu et al. 1988, 1994). The kinetic energy $E_{\rm out}$ of the 
outflow will generally be some fraction $\alpha$ of the binding 
energy of the star, i.e., 
$$E_{\rm out} = \alpha {GM_\ast^2 \over R_\ast} \, .  \eqno(2.2)$$
The natural time scale associated with stellar processes is 
the Kelvin-Helmholtz time scale.  We thus take the duration 
of the outflow (which is produced by a stellar process) to be a 
fraction $\beta$ of the Kelvin-Helmholtz time, i.e., 
$$\tau_{\rm out} = \beta {GM_\ast^2 \over R_\ast L_\ast} 
\, . \eqno(2.3)$$ 
Combining the above two equations, we thus reproduce the observational 
correlation that the mechanical outflow luminosity $L_{\rm out}$
$\equiv$ $E_{\rm out}/\tau_{\rm out}$ is roughly a constant fraction 
of the photon luminosity $L_\ast$ of the central source, i.e., 
$$L_{\rm out} = {\alpha \over \beta} \, L_\ast , \eqno(2.4)$$
where observations show that $\alpha/\beta$ $\sim 10^{-2}$ 
and that this correlation holds over several decades of $L_\ast$ 
(see Bally \& Lada 1983; Lada 1985; Levreault 1988; Edwards, Ray, 
\& Mundt 1993). Finally, if the winds roughly conserve energy 
while driving bipolar outflows, we obtain the result 
$$\dot M_w {G M_\ast \over R_\ast} = 
\epsilon {\alpha \over \beta} L_\ast , 
\eqno(2.5)$$
where $\epsilon$ is an additional efficiency parameter. 

The strength of the infall can be measured by the rate 
$\dot M_\ast$ at which matter falls {\it directly} onto the star.  
The total infall rate onto the central star/disk system is 
given by the collapse solution for an isothermal cloud core 
(Shu 1977).  This infall rate $\dot M$  for purely spherical 
infall takes the form 
$${\dot M} = m_0 a^3 / G \, , \eqno(2.6)$$
where $m_0$ = 0.975 is a dimensionless constant. For cloud cores
which are not isothermal, the mass infall rate can still be written 
in the form [2.6] with the sound speed $a$ taken to be the total 
effective sound speed (see equation [1.4]) and a different 
numerical constant (Adams et al. 1995). 

When rotation is present, not all of the material falls all 
the way in to the stellar surface (see Cassen \& Moosman 1981; 
Terebey, Shu, \& Cassen 1984).  The material with higher 
specific angular momentum collects in a circumstellar disk 
whose radius is roughly given by the centrifugal radius 
$$R_C \equiv {G^3 M^3 \Omega^2 \over 16 a^8} \, . \eqno(2.7)$$
When the stellar radius is small compared to the centrifugal 
radius, $R_\ast \ll R_C$, the direct infall rate $\dot M_\ast$ 
onto the star itself is a small fraction of the total mass infall 
rate and is given by 
$$\dot M_\ast = {R_\ast \over 2 R_C} {\dot M}
= {8m_0 R_\ast a^{11} \over G^4M^3\Omega^2} \, . 
\eqno(2.8)$$
The first equality is taken from equation [24] of Adams \& Shu 
(1986); the second equality follows from the expressions for 
$\dot M$ and $R_C$. 

In general, the stellar mass $M_\ast$ is only a fraction of 
the total mass $M$ that has collapsed to the central star/disk 
system at a given time; we write this condition in the form 
$$M_\ast = \gamma M \, .  \eqno(2.9)$$ 
Disk stability considerations (Adams, Ruden, \& Shu 1989; 
Shu et al. 1990) greatly limit the allowed range of the 
fraction $\gamma$.  We generally expect $\gamma \sim 2/3$; 
smaller values of $\gamma$ imply larger relative disk masses 
and hence systems that are gravitationally unstable. 

The combination of the above results implies that the final 
properties of the newly formed star are given by the following 
SEMF: 
$$L_\ast M_\ast^{2} = 8 m_0 \gamma^3 \delta 
{\beta \over \alpha \epsilon}  {a^{11} \over G^3\Omega^2} 
= \con \, {a^{11} \over G^3\Omega^2}  \, , \eqno(2.10)$$
where we have defined a new dimensionless parameter $\con$
in the second equality.  Under most circumstances, we expect that 
the parameters $\alpha$, $\beta$, $\gamma$, $\delta$, and $\epsilon$
can be estimated to a reasonable degree of accuracy.  For example, 
disk stability arguments suggest that $\gamma$ $\sim 2/3$ and
empirical estimates imply that $\beta/\alpha$ $\sim 10^2$.  The
parameter $\epsilon$ is close to unity, whereas the parameter $\delta$
has a value of a few.  We thus expect that the parameter $\con$ will
lie in the range $10^2 \le \con \le 10^3$.  In real star forming
environments, the parameters $\alpha$, $\beta$, $\dots$ do not have
exactly the same values for all cases. Instead, the values of these
parameters have a {\it distribution} which is determined by the
underlying physics of the problem.

Equation [2.10] provides us with a transformation between 
initial conditions (the sound speed $a$ and the rotation rate 
$\Omega$) and the final properties of the star (the luminosity 
$L_\ast$ and the mass $M_\ast$).  If we use ``typical'' values 
for present day clouds (e.g., $a$ = 0.35 km s$^{-1}$ and 
$\Omega \sim 3 \times 10^{-14}$ rad s$^{-1}$ $\sim$ 1 km s$^{-1}$ 
pc$^{-1}$) and the observed protostellar luminosities 
($L_\ast \sim 20 L_\odot$), we obtain stellar mass estimates 
$M_\ast$ $\sim$ 1 $M_\odot$, which is the typical mass of stars 
forming in regions with these properties.  In spite of its highly 
idealized nature, the SEMF [2.10] thus provides reasonable estimates 
for the mass $M_\ast$ as a function of initial conditions 
$(a, \Omega)$.  It is useful to write this transformation 
in dimensionless form 
$${\widetilde L} \, m^2 \, = \, 20 \con_3 \, a_{35}^{11} 
\, \Omega_1^{-2} \, , \eqno(2.11)$$
where we have defined 
${\widetilde L} \equiv L_\ast/(1 L_\odot)$, 
$m \equiv M_\ast/(1 M_\odot)$, 
$a_{35} \equiv a$/(0.35 km s$^{-1}$), 
$\Omega_1 \equiv \Omega$/(1 km s$^{-1}$ pc$^{-1}$), 
and finally $\con_3 \equiv \con / 10^3$. 

Notice that much of the uncertainty in this calculation has been 
encapsulated in the parameter $\con$, which should really 
be considered as a complicated function of all the 
environmental parameters.  Notice also that we have set 
the final mass of the star by the criterion that its outflow 
is sufficiently powerful to overwhelm the infall; in actuality, 
the star will continue to gain some mass, both from residual 
infall and from disk accretion, after this evolutionary state
has been reached.  This mass correction can be absorbed into 
the factor $\con$ in equation [2.10]. We have also characterized 
the initial conditions by only two physical variables ($a$ and 
$\Omega)$ whereas much more complicated initial states are 
possible.  Finally, we have ignored radiation pressure in 
this argument; for sufficiently massive stars ($M_\ast \ge$ 
7 $M_\odot$) radiation pressure will help the outflow reverse 
the infall (e.g., see Wolfire \& Cassinelli 1986, 1987;
Nakano 1989; Jijina \& Adams 1995). 

In order to evaluate the semi-empirical mass formula derived above, we
must determine the relationship between mass and luminosity for young
stellar objects. In general, the luminosity has many contributions
(Stahler, Shu, \& Taam 1980; Adams \& Shu 1986; Adams 1990; Palla \&
Stahler 1990, 1992).  For our present purposes, however, we can
simplify the picture considerably. The most important source of
luminosity for low mass objects is ultimately from infall; in other
words, infalling material falls through the gravitational potential
well of the star (and disk) and converts energy into photons.  This
luminosity can be written 
$$L_A = \eta {G M {\dot M} \over R_\ast} \, 
\approx 70 L_\odot \, \, \eta \, a_{35}^2 \, m \, , \eqno(2.12)$$
where $a_{35}$ and $m$ are the dimensionless sound speed and mass
as defined above. For the stellar radius, we have used the scaling 
relation $R_\ast = (3 \times 10^{11}$ cm )$\, a_{35}$, indicated by 
the stellar structure calculations of Stahler, Shu, and Taam (1980).
The efficiency parameter $\eta$ is the fraction of the total available
energy that is converted into photons. For spherical infall, all of
the material reaches the stellar surface and $\eta \approx 1$.  For
infall which includes rotation, some of the energy is stored in the
form of rotational and gravitational potential energy in the
circumstellar disk.  We generally expect $\eta \sim 1/2$.

In addition, the star can generate its own internal luminosity through
deuterium burning, gravitational contraction, and eventually hydrogen
burning.  This additional luminosity contribution is important for 
stars with masses larger than $\sim$ few $M_\odot$. For low mass stars
on the main sequence, the luminosity is a very sensitive function of
stellar mass, $L_\ast \sim M_\ast^4$.  For higher mass stars, the
mass/luminosity relation flattens to the form $L_\ast \sim M_\ast^2$
(and eventually flattens further to the form   $L_\ast \sim M_\ast$
for very high mass stars). 
For stars still gaining mass from infall, the internal luminosity 
contribution is somewhat different, but has been calculated for much 
of the relevant range of parameter space (Stahler 1983, 1988; 
Palla \& Stahler 1990, 1992; see also Fletcher \& Stahler 1994). 
For our present purposes, we use the following simple 
approximation for the internal luminosity 
$$L_{int} =  1 L_\odot \Bigl( {M_\ast \over 1 M_\odot} \Bigr)^4 
\, , \eqno(2.13)$$ 
which is roughly valid for the mass range 
$1 M_\odot \le M_\ast \le 10 M_\odot$. 

Putting both contributions together, we obtain the luminosity 
as a function of mass in dimensionless form:
$${\widetilde L} = 70 \, \eta \, a_{35}^2 \, m \,  
+ m^4 \, . \eqno(2.14)$$
Thus, at low masses, the luminosity is dominated by the contribution 
from infall and $\widetilde L$ is a linear function of $m$.  At 
higher masses ${\widetilde L} \sim m^4$ with the crossover point 
at $m \approx 3.3$ (for representative values of $\eta = 1/2$ 
and $a_{35}$ = 1).  At sufficiently high masses, the form [2.14] is 
no longer valid and ${\widetilde L} \sim 100 m^2$ for the range 
$10 \le m \le 100$. 

Finally, putting all of the above results together, 
we present the SEMF in dimensionless form: 

$$m = 0.66 [\con_3/\eta]^{1/3} \, a_{35}^3 \Omega_1^{-2/3} \, 
\qquad {\rm low} \, \, \, m \, , \eqno(2.15{\rm a})$$

$$m = 1.65 \, \con_3^{1/6} \, a_{35}^{11/6} \, \Omega_1^{-1/3} \, 
\qquad {\rm intermediate} \, \, \, m \, , \eqno(2.15{\rm b})$$

$$m = 0.67 \, \con_3^{1/4} \, a_{35}^{11/4} \, \Omega_1^{-1/2} \, 
\qquad {\rm high} \, \, \, m \, . \eqno(2.15{\rm c})$$

This SEMF provides a transformation between initial conditions and 
the final mass of the star.  One way to view this result is shown in
Figure 2.  Here, we assume that the effective sound speed $a_{35}$ 
and the rotation rate $\Omega_1$ are the two most important parameters
which determine the initial conditions.  We thus set all of the
remaining parameters to constant values such that $\con_3 = 1$ and we
take $\eta = 1/2$.  Figure 2 shows the resulting contours of constant
mass in the plane of initial conditions, i.e., the $(a_{35},
\Omega_1)$ plane.  The region in the far upper left corner of the
diagram corresponds to brown dwarfs, i.e., objects with masses less
than the hydrogen burning limit.  The region in the lower right corner
corresponds to stars that are too massive to be stable; the lower 
right part of the diagram also corresponds to initial conditions 
for which radiation pressure helps limit the stellar mass 
(see Jijina \& Adams 1995). In actual star forming regions, 
the parameters (in addition to $a_{35}$ and $\Omega_1$) which 
enter into the SEMF will have a distribution of values 
(roughly centered on the values assumed here).  As a result, 
the set of initial conditions which lead to a star of a given mass 
will be a band in the $(a_{35}, \Omega_1)$ plane instead of a line. 
Finally, for comparison, we note that previous authors have considered 
the mean density $n$ and the mass infall rate $\dot M$ as the two 
most important variables which determine stellar masses (see 
Figures 1 -- 3 of Nakano et al. 1995).

\bigskip 
\centerline{\bf 3. EMPIRICAL MODEL: THE INITIAL MASS FUNCTION} 
\centerline{\bf FOR CLUMPY MOLECULAR CLOUDS} 
\medskip 

In this section, we consider the limiting case in which the effective
sound speed is the only physical variable which determines stellar
masses.  Here, we use two observed scaling laws to determine the
distribution of the effective sound speed and hence the distribution
of initial conditions for star formation.  This result, in conjunction
with the SEMF derived in the previous section, produces a nearly
power-law IMF in reasonable agreement with observations.  We also
calculate the efficiency of star formation from this model.

\bigskip 
\centerline{\it 3.1 Observed Distributions of Initial Conditions} 
\medskip 

Observational work spanning many different star forming regions
and many different size scales suggests that the effective sound 
speed in molecular cloud cores (or clumps) obeys a simple 
scaling law (see, e.g., Larson 1981; Scalo 1987; Myers \& Fuller
1992).  For sufficiently large size scales ($r \sim$1 pc) and low
density $n < 10^4$ cm$^{-3}$, the observed linewidths $\Delta v$ in
cores have a substantial nonthermal component which scales with
density according to the law 
$$\Delta v \propto \rho^{-1/2} . \eqno(3.1)$$ 
Although this result was obtained from observational data, 
relations of this type can be calculated theoretically from the 
supposition that magnetohydrodynamic waves (e.g., Alfv{\' e}n 
and magnetoacoustic waves) are the source of the non-thermal 
motions (see Fatuzzo \& Adams 1993; McKee \& Zweibel 1995). 
Notice that this scaling law is valid over a finite range 
of densities; at sufficiently large densities the observed 
linewidths become equal to the thermal linewidths (in other 
words, the total linewidth $\Delta v$ does not vanish as 
$\rho \to \infty$ as implied by equation [3.1]). 

If this velocity $\Delta v$ is interpreted as a transport speed, 
then a ``turbulent'' or ``nonthermal'' component to the pressure 
can be derived (Lizano \& Shu 1989; Myers \& Fuller 1992) and 
has the form $P = P_0 \ln(\rho/\rho_0)$.  This ``turbulent'' 
equation of state also implies a scaling relation between the 
line-width $\Delta v$ and the mass $M_{\rm cl}$ of the clump. 
Using hydrostatic equilibrium arguments, we obtain the relation 
$$M_{\rm cl} \sim (\Delta v)^q \,  , \eqno(3.2{\rm a})$$
where $q \approx 4$ for the law given by equation [3.1].
Since we consider the linewidth $\Delta v$ to define the 
effective sound speed of the region, we can write this 
scaling relation in the form 
$$M_{\rm cl} = {\widehat M} a_{35}^q \,  , \eqno(3.2{\rm b})$$ 
where the mass scale $\widehat M$ $\sim$ 7 $M_\odot$ is 
determined by the normalization of the observed scaling law
(see, e.g., Larson 1981; Myers \& Fuller 1992). 

Given the above result, we must now determine the distribution of 
clump masses $M_{\rm cl}$.  Many groups have studied the observed 
clump mass spectrum of molecular clouds and have found 
nearly power-law forms, i.e.,
$${d N_{\rm cl} \over d M_{\rm cl} } \sim M_{\rm cl}^{-p}
\, , \eqno(3.3)$$
where the index of the power-law typically has the value 
$p \approx 3/2$ (e.g., see Scalo 1985; Lada, Bally, \& Stark 1991; 
Blitz 1993; Tatematsu et al. 1993).  This scaling relation must 
have a cutoff at both high mass (to keep the total mass of the 
cloud finite) and at low mass (to keep the total number of 
clumps finite). 

Although the distribution [3.3] was obtained from observations, such
distributions can, in principle, be calculated theoretically.  For
example, simple models which envisage clouds to be composed of a
collection of interacting clumps (Norman et al. 1995) can be used 
to derive clump mass distributions of the general form [3.3].  
We emphasize that much more theoretical work on this subject 
should be done. 

\bigskip 
\centerline{\it 3.2 A Simple Model for the IMF}
\medskip 

We can now piece together all of the above arguments to construct an
initial mass function. Here, we interpret the line-width $\Delta v$ as
the effective transport speed $a$ which determines the initial
condition for star formation.  We use the semi-empirical mass formula
of \S 2 to provide the transformation between the initial conditions
and the final stellar properties.  To start, we consider the simplest
case in which only the sound speed varies and the remaining parameters
of the SEMF are kept constant.  The relationship [3.2] between clump
mass and linewidth, in conjunction with the clump mass spectrum of
equation [3.3], determines the distribution of the effective sound
speed. Combining this distribution with the SEMF, we obtain an initial
mass function of the form
$$f = {d N \over d M_\ast} = {d N \over d M_{\rm cl} }
{d M_{\rm cl} \over d M_\ast} \sim M_\ast^{- \b}
\, , \eqno(3.4)$$
where we have assumed that only a single star forms in a given clump. 
The power-law index $\b$ of the distribution is given by 
$$\b = q (p-1)/\mu + 1 \approx 2/\mu + 1 \, , \eqno(3.5)$$ 
where $\mu$ is the scaling exponent which determines how 
the stellar mass varies with effective sound speed. As shown 
by equation [2.15], this index lies in the range 
11/6 $\le \mu \le $ 11/3.
Thus, this simple argument produces a power-law IMF with an
index in the range $\b = 1.6 - 2.1$.  This result compares
reasonably well with the observed power-law index of the IMF 
which has $\b \approx$ 2.35 (see Salpeter 1955). 

In Figure 3, we show the IMF calculated from this model. Here we 
use equations [3.2] and [3.3] to determine the distribution of the 
effective sound speed. We then use the SEMF in the form of equation
[2.11] and the mass/luminosity relationship [2.14].  The result is
shown as the dashed curve in Figure 3; as indicated by equation [3.4],
this distribution has a power-law index $\b \approx 1.6$ at low masses
and $\b \approx$ 2.1 at higher masses.  Also shown for comparison is
the analytic fit to the observed IMF (from Miller \& Scalo 1979).
Notice that the agreement between the theory and the observations is
quite reasonable, but is not exact.  We interpret this finding to mean
that this picture of the IMF is basically correct, but it still
incomplete.

The basic logic of this model can be summarized as follows.  
Molecular clouds produce a distribution of initial conditions for 
star formation.  In the simplest picture considered here, the clouds
produce a distribution of clump masses.  Because larger clumps have
larger effective sound speeds due to turbulence and other small-scale
physical processes, this distribution of clump masses implies a 
corresponding distribution of effective sound speeds, which represent 
the initial conditions for star formation.  We then use the idea that 
outflows help determine the final masses of forming stars to find a
transformation between the initial conditions and the final stellar
masses.  Using both this transformation and the set of initial 
conditions, we thereby obtain the IMF. 

\bigskip 
\centerline{\it 3.3 Efficiency of Star Formation} 
\medskip 

We can directly calculate the efficiency of star formation from 
this model of the IMF.  Here, we define the star formation 
efficiency $\eff$ to be the ratio of the mass in stars to the 
total cloud mass, i.e., 
$$\eff = \int_{M_1}^{M_2} {d N \over d M_{\rm cl} } M_\ast d M_{\rm cl}
\Bigg/
\int_{M_1}^{M_2} {d N \over d M_{\rm cl} } M_{\rm cl} d M_{\rm cl} 
\, , \eqno(3.6)$$ 
where $M_1$ and $M_2$ are the lower and upper cutoffs of the clump 
mass distribution [3.3].  Using the SEMF of the previous section
(for simplicity, we use only the low mass version [2.15a]) 
in conjunction with the scaling law of equation [3.2], we can 
evaluate this integral to obtain the efficiency 
$$\eff = 0.66 \, [\con_3/\eta]^{1/3} \, {2 - p \over 1 + 3/q - p} 
\, \, {\widehat m}^{-3/q} \, \, {m_2}^{3/q - 1} \, , \eqno(3.7)$$ 
where we have assumed that $M_1 \ll M_2$. 
We have also defined ${\widehat m} = {\widehat M} / (1 M_\odot)$ 
and $m_2 = M_2 / (1 M_\odot)$. 
If we use representative values, $\con_3 = 1$, $\eta = 1/2$, 
$p=3/2$, $q=4$, ${\widehat m}$ = 7, and $m_2 = 1000$ (e.g., see 
Williams, de Geus, \& Blitz 1994), we obtain a star formation 
efficiency $\eff \approx 0.07$. 

The star formation efficiency calculated here must be compared 
with the observed value for giant molecular clouds taken as a
whole.  This efficiency typically has a value of a few percent 
(e.g., Duerr, Imhoff, \& Lada 1982; Lada, Strom, \& Myers 1993). 
We conclude that this semi-empirical model produces 
a star formation efficiency in reasonable agreement with 
observations. 

For completeness, we calculate the efficiency of star formation 
for an individual clump. This efficiency $\eff_\ast$ in 
the low mass regime is then given by 
$$\eff_\ast = {M_\ast \over M_{\rm cl} } = 
{0.66 \over {\widehat m} } 
[\con_3/\eta]^{1/3} \, a_{35}^{3-q} \approx 
0.12 \, a_{35}^{-1} \, , \eqno(3.8)$$
where the final approximate equality was obtained using
the representative values $\con_3 = 1$, $\eta = 1/2$, 
$q=4$, and ${\widehat m}$ = 7.  A similar formula can be
derived for the regimes of intermediate and high mass stars. 

In this model, we have assumed that only a single star forms within a
given molecular clump.  In many cases (see Zinnecker et al. 1993; Lada
et al. 1993; Hillenbrand et al. 1993), a cluster of stars forms within
a single clump and the star formation efficiency of the individual
clump can be much higher, $\eff_{\rm clust} \sim 0.2 - 0.5$.  This
present model does not take cluster formation into account, although
this issue is important and should be addressed in future studies.

\newpage
\centerline{\bf 4. RANDOM MODEL: THE INITIAL MASS FUNCTION}
\nobreak
\centerline{\bf AS A RESULT OF THE CENTRAL LIMIT THEOREM} 
\nobreak
\medskip 

In this section, we consider the limit in which a large number 
of physical variables is required to determine stellar masses. 
We thus adopt a statistical approach to the calculation of
the IMF.  We start with the semi-empirical mass formula of \S 2 and
consider it to be a product of random variables.  In the limit that
the number $n$ of random variables is large ($n \to \infty$) and the
variables are completely independent, the IMF approaches a 
log-normal distribution.  This result is a direct consequence of the
central limit theorem (see, e.g., Richtmyer 1978; Parzen 1960). 
For the more realistic case of a finite number of 
not-completely-independent variables, we must study how the
resulting distribution differs from a log-normal distribution (see 
also \S 5).  The idea of using the central limit theorem to obtain 
a log-normal distribution has been discussed previously in models
where the stellar masses are determined by fragmentation (Larson 1973;
Elmegreen \& Mathieu 1983; Zinnecker 1984, 1985).  In this paper, 
we adopt a different approach using the SEMF of \S 2 as the 
starting point for our calculation.  We also note that the 
number $n$ of physical variables is actually finite and hence 
departures of the IMF from a log-normal form are important. 

The semi-empirical mass formula can be written in the general 
form of a product of variables
$$M_\ast = \prod_{j=1}^n \alpha_j \, , \eqno(4.1)$$
where the $\alpha_j$ represent the various quantities on the 
right hand side of equation [2.10], i.e., $\alpha$, $\beta$, 
$\gamma$, $\delta$, $\epsilon$, $a$, and $\Omega$ (taken to 
the appropriate powers). In this present discussion, we regard 
these quantities $\alpha_j$ as a collection of $n$ random variables. 
Thus, by taking the logarithm of this equation, we find that the 
logarithm of the mass is a sum of random variables, 
$$\ln M_\ast = \sum_{j=1}^n \ln \alpha_j 
+ {\sl constant} \, , \eqno(4.2)$$
where the constant term includes all the quantities in the SEMF 
that are truly constant (e.g., the gravitational constant $G$). 

The stellar mass is thus determined by a composite random variable
that is given by the sum of random variables.  The central limit
theorem shows that the distribution for the composite variable 
always approaches a normal (gaussian) distribution as the number 
$n$ of variables approaches infinity.  In order to use this result, 
we must redefine the basic variables $\ln \alpha_j$ so that the new 
variables $\xi_j$ have zero mean, i.e., 
$$\int_{-\infty}^\infty \, \xi_j f_j (\xi_j) d\xi_j = 0 \, , \eqno(4.3)$$
where $f_j$ is the probability density of the $jth$ variable. 
The distribution $f_j$ is, in general, not a normal (gaussian) 
distribution. These new variables $\xi_j$ are related to the 
old variables $\alpha_j$ through the relation 
$$\xi_j \equiv  \, \ln\alpha_j - 
\langle \ln\alpha_j \rangle \equiv  
\ln[\alpha_j / \alphabar_j ] \, , \eqno(4.4)$$
where angular brackets represent averages.  Keep in mind that 
the averages are taken over the logarithms of the $\alpha_j$ and 
not over the variables $\alpha_j$ themselves, i.e., 
$$\ln \alphabar_j = \langle \ln\alpha_j \rangle = 
\int_{-\infty}^\infty \ln\alpha_j \, f_j (\ln\alpha_j) \,
d\ln\alpha_j \, . \eqno(4.5)$$
Similarly, the distributions $f_j$ are the distributions of 
$\ln \alpha_j$ and not the distributions of $\alpha_j$. 
Each of the rescaled variables $\xi_j$ has a variance 
$\sigma_j$ given by 
$$\sigma_j^2 \, = \, \int_{-\infty}^\infty \, \xi_j^2 
f_j (\xi_j) d\xi_j \,  . \eqno(4.6)$$ 

Next we construct a composite random variable $\zeta$, 
defined by 
$$\zeta \equiv \sum_{j=1}^n  \xi_j \, = \sum_{j=1}^n  
\ln[\alpha_j / \alphabar_j ] \, .  \eqno(4.7)$$
In terms of this new variable, the semi-empirical 
mass formula becomes 
$$M_\ast = M_C \, \, {\rm e}^\zeta \, , \eqno(4.8)$$
where $M_C$ is a characteristic mass scale.  The distribution of 
stellar masses (the IMF) is thus determined by the distribution 
of the composite variable $\zeta$.  The mass scale $M_C$ is 
determined by the mean values of the logarithms of the original 
variables $\alpha_j$, i.e., 
$$M_C \equiv \prod_{j=1}^n \exp [ \langle \ln \alpha_j \rangle ] 
\equiv \prod_{j=1}^n \alphabar_j \, , \eqno(4.9)$$
where we have defined $\alphabar_j$ = 
$\exp [ \langle \ln \alpha_j \rangle ]$. 

As we find below, the variance $\sigbar$ of the composite 
variable $\zeta$ essentially determines the width of the stellar 
mass distribution and is thus of fundamental importance for this 
present discussion.  The variance is given by 
$$\sigbar^2 = \sum_{j=1}^n \sum_{k=1}^n \int_{-\infty}^\infty 
\int_{-\infty}^\infty d\xi_j d\xi_k \, \xi_j \, \xi_k \, 
f(\xi_j, \xi_k) \, , \eqno(4.10)$$ 
where $f(\xi_j, \xi_k)$ is the joint probability distribution of 
the variables $\xi_j$ and $\xi_k$.  {\it If the variables are 
statistically independent}, then the joint probability is the 
product of the individual probability distributions and the 
integral in equation [4.10] can be separated.  Using the fact 
that each of the variables $\xi_j$ has zero mean (equation [4.3]) 
and the definition [4.6], we thus obtain the total variance, 
$$\sigbar^2 = \sum_{j=1}^n \sigma_j^2 \, . \eqno(4.11)$$
We can now define a new random variable 
$$\tz \equiv {\zeta \over \sigbar} \, , \eqno(4.12)$$
which has zero mean and unit variance.  
The central limit theorem tell us that the distribution of the 
composite random variable $\tz$ approaches a normal distribution 
in the limit $n \to \infty$, i.e., 
$$f(\tz) \to {\cal N} {\rm e}^{-\tz^2/2} \, , \eqno(4.13)$$
where ${\cal N}$ = $1/\sqrt{2 \pi}$ is a normalization constant. 
This result is {\it independent} of the initial distributions 
$f_j$. 

The mass of the star (from equation [4.1]) is related to 
this new variable $\tz$ through the relation 
$$\ln M_\ast = \ln M_C + \sigbar \tz \, , \eqno(4.14)$$
where the variable $\tz$ now has a known (gaussian) distribution. 
Combining equations [4.13] and [4.14], we can write the 
distribution $f$ of stellar masses in the form 
$$\ln f (\ln m) = A - {1 \over 2 \sigbar^2} 
\Bigl\{ \ln \bigl[ m / m_C \bigr] \Bigr\}^2 \, , \eqno(4.15)$$
where $A$ is a constant and where we have defined 
$m \equiv M_\ast/(1 M_\odot)$ and 
$m_C \equiv M_C / (1 M_\odot)$.
Notice that the constant $A$ just sets the overall 
normalization of the distribution. The shape of the 
distribution is thus completely determined by the 
mass scale $m_C$ and the total variance $\sigbar$. 

In this limit, the distribution of stellar masses has 
exactly the same form as the Miller/Scalo approximation of 
equation [1.3].  The three constants $a_1$, $a_2$, and 
$a_3$ of the Miller/Scalo law are related to the physical 
variables $A$, $m_C$, and $\sigbar$ through the relations 

$$A = a_0 \ln 10 + {a_1^2 \ln 10 \over 4 a_2} 
\approx 4.65 \, , \eqno(4.16{\rm a})$$

$$\sigbar^2 = {\ln 10 \over 2 a_2 } \approx 2.45 \, , 
\eqno(4.16{\rm b})$$

$$\ln m_C = - {a_1 \ln 10 \over 2 a_2} \approx -2.35
\, . \eqno(4.16{\rm c})$$
Thus, the ``characteristic mass scale'' of the distribution 
is $m_C \approx 0.095$.

In this limit, where the SEMF involves a large number
of statistically independent variables, we obtain a ``pure''
log-normal distribution.  Since the observed IMF can be 
approximately fit by a log-normal distribution (see Figure 1), 
this model of the IMF is in reasonable agreement with observations.  
In this limit, the only relevant parameters are the total width 
of the distribution (determined by $\sigbar$) and the center of 
the distribution (determined by $m_C$).
The product of the mean values of all the of relevant variables
combine to determine the mass scale $m_C$ of the distribution (see
equation [4.9]).  Similarly, the widths of all of the original
variables combine to determine the total width $\sigbar$ of the 
final distribution (see equation [4.11]). 

We thus obtain an important consistency check on this model of 
the IMF: The total width $\sigbar$ and the characteristic mass 
scale $m_C$ can be calculated and compared with the values required
to fit the observed IMF (see equation [4.16]).  The quantities 
$\sigbar$ and $m_C$ are determined by the distributions of all 
of the physical variables in the problem.  In a complete theory, 
we could calculate these initial distributions from {\it a priori} 
considerations.  In the absence of a complete theory, however, we 
can use observations of the physical variables to estimate their 
distributions and hence calculate $\sigbar$ and $m_C$. Such a 
calculation is performed in Appendix B. Using estimates of the 
distributions of the observed physical variables $a_{\rm eff}$, 
$\Omega$, etc., we obtain values $\sigbar \approx$ 1.8 
and $m_C \approx$ 0.25.  Although these values are somewhat 
higher than the values required to fit the Miller/Scalo IMF 
(see equation [4.16]), we consider the agreement to be quite 
good, given the crudeness of the calculation.  

\bigskip 
\centerline{\bf 5. INTERMEDIATE EXAMPLES AND APPLICATIONS} 
\medskip 

The previous discussion has considered the two limiting cases in which
the number $n$ of physical variables that determine stellar masses is
either one (\S 3) or infinite (\S 4).  In the first case, we obtain a
nearly power-law IMF; in the second case, we obtain a log-normal IMF.
In realistic star forming regions, however, we must consider the
intermediate cases with $1 < n < \infty$.  In particular, we must
determine the form of the composite distribution (the IMF) for these 
intermediate cases. In general, the answer depends on the
distributions of the initial variables. In this section, we explicitly
calculate theoretical IMFs for several different cases with various
distributions of initial conditions. We also consider the problem
of binary companions.  Although the theory discussed thus far only
applies directly to the formation of single stars, we can show (\S
5.3) that the inclusion of binary companions does not greatly change
the resulting IMF.

\bigskip 
\centerline{\it 5.1 Uniform Distributions} 
\medskip 

In this subsection, we consider the simple case of 
$n$ fundamental variables $\alpha_j$, each with the 
same distribution $f_j (\ln \alpha_j)$.  We also consider 
the simplest type of distribution in which $\xi_j = \ln \alpha_j$
is uniformly distributed in an interval $[-w, w]$, i.e., 
$$\eqalign{ f_j (\xi_j) = & \, {1 \over 2 w} \qquad {\rm for} 
\, -w \le \xi_j \le w \, , \cr 
= & \, 0 \qquad \quad {\rm otherwise} \, . } \eqno(5.1)$$
The variance $\sigma_j$ of each individual variable is related 
to the width $w$ of the interval by the expression 
$$\sigma_j^2 = w^2 / 3 \, . \eqno(5.2)$$
For a given number $n$ of variables, we can thus obtain the 
required total width $\sigbar$ of the distribution by taking 
$$w^2 = {3 \over n} \sigbar^2 \, , \eqno(5.3)$$ 
where we use equation [4.16b] to set the value of $\sigbar$. 
We also use equation [4.16c] to set $m_C$ and hence the center
of the distribution. 

The resulting distribution (IMF) is shown in Figure 4 for the case
$n=10$.  To obtain this result, we have used a random number 
generator to produce the SEMF variables (distributed according to 
equation [5.1]) and have calculated a million ($10^6$) realizations 
of the mass.  Notice that these random variables {\it do not} have 
a gaussian distribution, but the sum of random variables comes rather 
close to a log-normal distribution and hence reproduces the 
Miller/Scalo IMF quite well. In other words, $n=10$ is ``close 
enough to $\infty$'' for the central limit theorem to apply and 
hence for the distribution of $\zeta$ to be nearly gaussian, i.e., 
for the IMF to have nearly a log-normal form. 

Next, we would like to determine how the composite distribution
changes with the number $n$ of fundamental variables $\xi_j$.
Although the result depends on the initial distributions $f_j$ of the
variables, we can get some feeling for this problem by using the
uniform distribution [5.1] and varying the number $n$.  The result is
shown in Figure 5 for the cases $n$=1, 2, 3, and 5.  The mass scale
$m_C$ has been set to correspond to that of the Miller/Scalo IMF
(shown as the solid curve).  Notice that the composite distribution 
converges toward the log-normal limit rather rapidly with increasing 
values of $n$.  As expected, the largest departures are for high
masses, i.e., for the tail of the distribution.

\bigskip 
\centerline{\it 5.2 Power-law Distributions} 
\medskip 

In this subsection, we explicitly consider the case in which all 
of the physical variables appearing in the SEMF have power-law 
distributions.  This case is expected to be a reasonable 
approximation for many astrophysical systems. 

We consider each fundamental variable to 
have a distribution of the form 
$$ {dN \over d \alpha_j} = C \alpha_j^{-p} \, , 
\eqno(5.4)$$
where $C$ is the normalization constant. 
If the power-law index $p > 1$, as we assume here, then we 
must specify the lowest value of the parameter $\alpha_{0j}$, 
i.e., we introduce a lower cutoff but not an upper cutoff. 
Next, we write the distribution in terms of the logarithmic 
variable $\xi_j$, which is defined by 
$$\xi_j \equiv (p - 1) \ln (\alpha_j/\alpha_{0j}) - 1
\, . \eqno(5.5)$$
The corresponding distribution $f_j$ is then given by 
$$f_j = {\rm e}^{- (\xi_j + 1)} \, \, , \eqno(5.6)$$
i.e., we obtain an exponential distribution. 
Notice that this new variable $\xi_j$ 
has zero mean and unit variance. 

For this case, we can explicitly calculate the composite 
distribution, and hence the IMF, from the initial distributions. 
Here we employ standard methods from probability theory (e.g., 
Richtmyer 1978; Parzen 1960).  The Fourier transform $\chi_j$ 
of the distribution $f_j$ ($\chi_j$ is generally known as the 
characteristic function of the variable $\xi_j$) is defined by 
the usual integral 
$$\chi_j (\lambda) \equiv \int_{-1}^\infty d\xi \, 
{\rm e}^{-i \lambda \xi} {\rm e}^{-(\xi+1)} \, = 
{ {\rm e}^{i \lambda} \over 1 + i \lambda} \, ,  \eqno(5.7)$$
where we have evaluated the integral to obtain the second equality. 

Next, we define a composite variable $\tz$ according to 
$$\tz \equiv {1 \over \sqrt{n} } \, 
\sum_{j=1}^n \, \xi_j \, \, , \eqno(5.8)$$
which also has zero mean and unit variance. 
The Fourier transform $\chi_\zeta$ of the composite 
distribution is then the product of the Fourier transforms 
of the individual distributions evaluated at $\lambda/\sqrt{n}$. 
Thus, the composite transform $\chi_\zeta$ is given by 
$$\chi_\zeta (\lambda) = 
\Bigl[ \chi_j (\lambda/\sqrt{n}) \Bigr]^n = 
\Bigl[ { {\rm e}^{i \lambda/\sqrt{n} } \over 
1 + i \lambda/\sqrt{n} } \Bigr]^n \, . \eqno(5.9)$$
To obtain the composite distribution itself $f_n(\tz)$, 
we simply take the inverse Fourier transform of 
equation [5.9]. We thus obtain  
$$f_n (\tz) = { \sqrt{n} \over (n-1)!} 
\Bigl\{ n + \sqrt{n} \tz \Bigr\}^{n-1} 
{\rm e}^{- ( n + \sqrt{n} \tz ) } \, , \eqno(5.10)$$
i.e., we obtain a {\it gamma distribution} for the composite 
variable $\tz$.  It is straightforward to show that in the 
limit $n \to \infty$, the distribution [5.10] approaches a 
log-normal form (as the central limit theorem requires).  
However, for the special case of power-law initial distributions, 
we have an exact form for the distribution for intermediate 
cases (i.e., for finite values of $n$). 

The total variance $\sigbar$ of the composite distribution 
depends on the power-law indices $p$ and the number of 
variables through the relation 
$$\sigbar^2 = {n \over (p-1)^2} \, . \eqno(5.11)$$
Similarly, the central mass scale $m_C$ of the distribution 
is given by 
$$m_C = {n \over p-1} \, + \, \sum_{j=1}^n \ln \alpha_{0j} 
\, . \eqno(5.12)$$

In Figure 6, we show the composite distributions $f_n$ for various
values of $n$.  Also shown for comparison is the log-normal form 
which corresponds to the limit $n \to \infty$.  All distributions 
shown here have the same total variance $\sigbar$ and characteristic 
mass scale $m_C$ as given by equation [4.16].  Notice that the 
convergence to the log-normal form is much slower than for the 
case of uniform distributions considered in the previous section.  
In particular, for sufficiently large masses, the IMF falls off 
like a decaying exponential in the variable $\ln m$ (a power-law 
in the variable $m$) instead of like a gaussian.  We note that 
the observed IMF seems to have a power-law tail at high masses, 
although the exact slope is somewhat uncertain (see, e.g., 
Massey et al. 1995 for a good discussion of this issue). 

\bigskip 
\centerline{\it 5.3 The Effects of Binary Companions} 
\medskip

In this section, we address the issue of binary companions.  
Thus far, the discussion of this paper has focused on the formation 
of a single star. On the other hand, most stars live within binary 
systems (see, e.g., Abt \& Levy 1976; Abt 1983; Duquennoy \& Mayor 
1991; Bodenheimer, Ruzmaikina, \& Mathieu 1993).  Several different 
mechanisms for producing binary companions have been proposed,
including capture via star/disk interactions (Clarke \& Pringle 1991),
formation from gravitational instabilities in circumstellar disks
(Adams, Ruden, \& Shu 1989), and fragmentation during protostellar
collapse (Boss 1992; Bonnell \& Bastien 1992).  Unfortunately,
however, a complete theory of binary formation has not yet been
obtained.  As a result, we must once again proceed in a 
semi-empirical manner. 

The SEMF of \S 2 can be interpreted as providing the final 
mass $M_{P \ast}$ of the {\it primary} star in a binary system. 
If we can write this SEMF in the form of equation [4.1], 
then the mass of the secondary $M_{S \ast}$ can also 
be written in the general form of a product of variables, i.e., 
$$M_{S \ast} = \alpha_S \prod_{j=1}^n \alpha_j \, , \eqno(5.13)$$
where $\alpha_S$ = $M_{S \ast}/M_{P \ast}$ is the mass ratio of 
the binary system.  Note that $\alpha_S \le 1$ by definition.
A complete theory of binary formation would give us a theoretical
estimate of the distribution of the mass ratio $\alpha_S$.  As
mentioned above, however, we do not yet have a complete theory 
of binary formation. As a result, we must use observations to measure 
and/or constrain the distribution of the mass ratio.  For a given 
distribution of $\alpha_S$, we can determine the consequences 
for the IMF using the framework developed in this paper. 

We first note that if the distribution of masses for the primary star
has a known form (e.g., the log-normal form of \S 4), then the
distribution of the secondary masses has nearly the same form. 
This result follows directly from equation [5.13] for the SEMF and
holds for any $\alpha_S$ distribution that is not overly pathological
(this statement can be made mathematically more precise -- see
Richtmyer 1978; Parzen 1960).  Furthermore, the width $\sigbar_S$ 
of this secondary distribution is directly determined from the width 
$\sigbar_P$ of the primary distribution through the relation 
$$\sigbar_S^2 = \sigbar_P^2 + \sigma_{\alpha S}^2 \, , 
\eqno(5.14)$$
where $\sigma_{\alpha S}$ is the variance of the mass ratio 
distribution.  In obtaining this result, we have assumed that 
the distribution of the mass ratio $\alpha_S$ is independent 
of the other variables in the problem. If the ratio $\alpha_S$
is not completely independent, then the expression [5.14] provides
an upper limit to the variance of the secondary distribution. 
Notice that the width of the secondary distribution is always 
wider than that of the primary distribution. 
Similarly, the characteristic mass scale $m_{C S}$ of the 
secondary distribution is given by 
$$m_{C S} / m_{C P} = {\rm e}^{\langle \ln \alpha_S \rangle} 
\, , \eqno(5.15)$$ 
where $m_{C P}$ is the mass scale of the primary distribution 
and where angular brackets denote averages over the distribution.  

As a reference point, we consider the simplest case in which the 
mass ratio $\alpha_S$ is uniformly (randomly) distributed over 
the interval $[0,1]$.  For this case, $\sigma_{\alpha S}$ = 1 
and $\langle \ln \alpha_S \rangle = -1$.  Although the observed  
distribution of mass ratios is not completely uniform, these values 
for $\sigma_{\alpha S}$ and $\langle \ln \alpha_S \rangle$ 
represent reasonable estimates (see, e.g., Duquennoy \& Mayor 1991
for a more detailed discussion). 

The total mass distribution, including both primary and secondary 
stars, is the sum of the two individual distributions.  For the 
sake of definiteness, we use the large $n$ limit for which both 
distributions obtain a nearly log-normal form.  In this case, 
we obtain the total distribution in the form 
$$f = {\cal N} \Bigl\{ {\rm e}^{-\zeta^2/2} + {\cal F} 
{\rm e}^{-(\zeta + \zeta_0)^2/2B^2}
\Bigr\} \, , \eqno(5.16)$$
where $\cal N$ is the normalization constant,
$\cal F$ is the binary fraction, and $\zeta = \ln m/m_{CP}$ 
is the composite variable for the primary mass distribution 
(see equations [4.7] and [4.8]).  We have also defined a 
parameter $B$ which represents the ratio of the widths of 
the secondary and primary mass distributions, 
$$B \equiv \bigl[ 1 + \sigma_{\alpha S}^2 / \sigbar_P^2 
\bigr]^{1/2} \, , \eqno(5.17)$$
and a parameter $\zeta_0$ which determines the difference 
in the centers of the two distributions, 
$$\zeta_0 \equiv - { \langle \ln \alpha_S \rangle 
\over \sigbar_P} \, . \eqno(5.18)$$ 
In the limit $B \to 1$ and $\zeta_0 \to 0$, we recover the 
primary distribution for the IMF.  For relatively small departures 
from this limit, the joint distribution $f$ is nearly the same as 
the primary distribution.  For the reference case of a uniform 
distribution of the mass ratio $\alpha_S$, we obtain values 
$B \approx$ 1.2 and $\zeta_0 \approx$ 0.64.

We can quantify the difference between the joint distribution 
$f$ and the original distribution $f_0$ for the masses of 
the primary stars.  We define an error functional $E_R$, 
$$E_R [f] \equiv 2 \sqrt{\pi} \, \int_{-\infty}^\infty \, 
\big| f_0 - f \big|^2 \, d\zeta \, , \eqno(5.19)$$ 
where we have normalized the integral such that $E_R [f=0]$ = 1. 
The size of the error estimate $E_R$ is thus a measure of how 
the distribution $f$ differs from the original distribution 
$f_0$.  For the joint distribution given by equation [5.16], 
we evaluate this functional to obtain  
$$E_R = {B^2 {\cal F}^2 \over (1 + B {\cal F})^2} 
\Bigl\{ 1 + B^{-1} - 2 \sqrt{2} (1+B^2)^{-1/2} \, 
{\rm e}^{-\zeta_0^2/2(1 + B^2)} \, \Bigr\} \, . \eqno(5.20)$$ 
Using the uniform distribution to estimate $\zeta_0$ and $B$, 
we find $E_R \approx 0.04$. We thus conclude that the effects 
of binary companions on the IMF are not overly large for the 
paradigm considered in this paper. 

In some sense, the error estimate obtained above is overly 
conservative because it includes the differences between the two 
distributions over the entire mass range $0 \le m \le \infty$. 
If we normalize the two distributions to unity for a mass of 
1.0 $M_\odot$ and only consider the expected mass range for 
stars, the difference is much smaller.  This result is 
shown in Figure 7.  The solid curve shows the primary 
distribution $f_0$ with width and mass scale consistent 
with the Miller/Scalo estimate. The dashed curve shows the 
effect of adding a distribution of binary companions 
according to equation [5.16] with $\zeta_0$ = 0.64, 
$B$ = 1.2, and binary fraction $\cal F$ = 0.75. 

\bigskip 
\bigskip 
\centerline{\bf 6. SUMMARY AND DISCUSSION} 
\medskip 

\bigskip 
\centerline{\it 6.1 Summary of Results}  
\medskip 

In this paper, we have presented models of the initial mass 
function using the idea that stars, in part, determine their 
own masses through the action of stellar winds and outflows. 
Our results can be summarized as follows:

\item{[1]} We have presented a semi-empirical mass formula (SEMF) 
for forming stars (see equation [2.10]).  This result determines 
the transformation between the initial conditions for star formation 
and the final masses of forming stars and uses the idea that stars 
determine their own masses through the action of stellar winds and
outflows (Figure 2; see also SLA).

\item{[2]} We have presented an empirical model for the IMF in the
limit that the effective sound speed is the most important physical
variable which determines stellar masses.  In this limit, the spectrum
of initial conditions for star formation is given by the combination
of the observed clump mass distribution [3.3] and the relationships
between clump mass, density, and linewidth (equations [3.1] and
[3.2]).  This distribution in conjunction with the SEMF produces a
nearly power-law distribution of masses of forming stars (see 
Figure 3). This theoretical distribution is in reasonable agreement 
with the observed IMF. 

\item{[3]} The empirical model of the IMF also allows us to 
estimate the overall star formation efficiency $\eff \sim 0.07$. 
This calculated efficiency (see equation [3.7]) is in reasonable 
agreement with observations. 

\item{[4]} We have studied the IMF in the limit where a large 
number $n$ of physical parameters play a role in determining stellar
masses. The central limit theorem shows that for any SEMF which can 
be written as the product of parameters (as in equation [4.1]) the 
resulting distribution of stellar masses (the IMF) approaches a
log-normal distribution as the number of parameters $n \to \infty$.
Since the observed IMF is crudely given by a log-normal distribution, 
this theory is in reasonable agreement with observations as long as 
the number $n$ of parameters which characterize star forming 
environments is sufficiently large. 

\item{[5]} When the central limit theorem applies (item [4]), the
resulting IMF is specified by two quantities: the total width
$\sigbar$ of the distribution and the characteristic mass scale $m_C$.
For a given SEMF, both of these quantities can be {\it calculated} 
from the original parameters in the problem.  The total width
$\sigbar$ is determined by the quadrature sum of the variances of the
distributions of all of the input parameters of the problem (see
equation [4.11]). The mass scale $m_C$ is determined by the average 
of the logarithms of the input parameters (see equation [4.9]). For
the SEMF of \S 2, the values of $\sigbar$ and $m_C$ estimated from
observed distributions of the input parameters are in basic agreement
with those required to fit the observed IMF (see Appendix B).

\item{[6]} We have studied the IMF resulting from the physically 
realistic case of intermediate numbers of variables ($1 < n < \infty$). 
In this case, the exact form for the IMF depends on the distributions
of the original physical parameters of the problem.  When these input
parameters have uniform (flat) distributions, the convergence of the
IMF to a log-normal form is quite rapid (Figures 4 and 5). For the
case of power-law distributions of the initial variables, the
convergence is much slower and the IMF retains a power-law tail at
high masses (Figure 6).

\item{[7]} We have briefly considered the effects of binary 
companions on the theory of the IMF presented in this paper.  
We show that the inclusion of binaries does not greatly change 
the resulting IMF; in addition, the manner in which binaries 
change the IMF can be directly calculated 
(see \S 5.3 and Figure 7). 

\item{[8]} The combination of all of these results demonstrates 
the consistency of the hypothesis that winds and outflows 
help determine the masses of forming stars by limiting the 
infall. 

\bigskip 
\centerline{\it 6.2 Discussion} 
\medskip 

At this point, we must carefully assess what we have calculated and
what we have not.  We have studied the idea that stars, in part,
determine their own masses through the action of strong stellar winds
and outflows.  The question thus becomes: Have we proven this
conjecture or have we presented mere speculation?  The results of this
paper represent much more than the latter, but unfortunately much less
than the former. We have demonstrated the {\it plausibility} of our
hypothesis, but we have certainly fallen short of a full proof.  In
this work, we have demonstrated that the idea of stars determining
their own masses through the action of stellar outflows is in fact
compatible with the observed distribution of stellar masses.

Perhaps the most important result of this work is that it provides a
framework to approach the calculation of the IMF.  In this framework,
the calculation of the IMF involves the two steps outlined in the
introduction: (1) the selection of initial conditions and (2) the
transformation between a given set of initial conditions and the final
mass of the star (see also Zinnecker 1989, 1990). Although treatments 
of both of these steps have been given here, the calculation of each
of these steps can be refined considerably (see \S 6.5).

The semi-empirical model (\S 3) and the random model (\S 4), as
presented here, are opposite limits of the same underlying problem.
In the empirical limit, the distribution of sound speed {\it is} the
distribution of initial conditions.  In the opposite limit of a random
model with a large number of independent variables, the total
distribution of initial conditions is independent of the distribution
of the sound speed except for its contribution to the overall width of
the distribution.  The actual physical case lies between these two
extremes, i.e., the IMF should be determined by {\it several}
variables (say, $n=3 - 10$) which are not completely independent.
However, even in the limit of a single variable -- the effective sound
speed -- we obtain a nearly-power-law distribution which is reasonably 
close to the observed IMF. Furthermore, relatively few independent
variables are necessary to ``round out'' the distribution to be 
even closer to the observed distribution (see Figure 5). 

Next, we must consider the issue of uniqueness.  In the limit that a
large number of physical variables play a role in the star formation
process, the central limit theorem implies that the resulting
composite distribution (the IMF) always approaches a log-normal form.
As a result, many different theories can, in principle, predict very
nearly the same IMF. The best way to discriminate between competing
theories is thus to look for the deviations of the theoretically
predicted IMF from a pure log-normal form.  As a general rule, 
the IMF will deviate most from log-normal at the {\it tails} of the 
distribution, i.e., at the low-mass and high-mass ends.  It is 
thus crucial to obtain tight observational constraints on the 
IMF at both high and low masses.  Unfortunately, however, the 
IMF is notoriously difficult to determine at both the high mass 
end (Massey et al. 1995) and the low mass end (Tinney 1995).
This issue represents a challenge for the future. 

\bigskip 
\centerline{\it 6.3 Implications of the Theory} 
\medskip 

The picture of the IMF promoted in this paper can be tested, 
or at least highly constrained, by observations.  In addition, 
this theory makes several preliminary predictions, which we 
discuss below. 

The first issue is that of the semi-empirical mass formula. 
This result provides a transformation between the initial 
conditions for star formation and the final stellar properties. 
In this picture, the final stellar mass is most sensitive to 
the total effective sound speed and the relation has the form 
$M_\ast \sim a^\mu$ with $2 \le \mu \le 3$.  This theoretical 
result is in good agreement with the observed correlation 
between the mass $M_{\ast max}$ of the largest star in a region 
and the observed line-width $(\Delta v)$ in that region 
(Myers \& Fuller 1993); the observed relation can be written  
$$M_{\ast max} \propto (\Delta v)^{2.38 \pm 0.17} \, , $$
and is valid for the mass range $0.1 < m < 30$. 
The theoretical relation $M_\ast \sim a^\mu$ is a direct result 
of the hypothesis that stars help determine their masses through 
the action of stellar winds and outflows. Furthermore, the SEMF 
has the same general form for a variety of cases and hence this 
result is fairly robust (see Appendix A). 

We have shown that as the number of variables in the SEMF increases,
the form of the IMF approaches a log-normal distribution, provided 
only that many different variables play a role in the SEMF.
The degree to which an exact log-normal distribution is obtained is
illustrated by Figures 3 -- 7.  Thus, one prediction of this theory is
that the IMF should have (nearly) a log-normal form for {\it any} star
forming environment.  Furthermore, when the IMF deviates from a pure  
log-normal distribution, it is expected to have a power-law tail at
high masses as illustrated by Figure 6. 

The mass distribution can be characterized by two parameters: the mass
scale $m_C$ and the total width $\sigbar$ of the distribution.  For a
given SEMF, the values of these parameters $\sigbar$ and $m_C$ can be
calculated from the distributions of the original variables in the
problem (see equations [4.9] and [4.11]).  Although the underlying
distributions might not be known exactly, the width and central values
of the distributions may be estimated.  Such estimates, in conjunction
with the results of this paper, may be useful in determining the IMF
for models of galaxy formation, cooling flows, and other astrophysical
systems.

\bigskip 
\centerline{\it 6.4 The Low-Mass End of the IMF and Brown Dwarfs} 
\medskip 

Another important issue is the lower mass cutoff for stars.  The
search for brown dwarfs has interested astronomers for many decades,
both as a limiting case of stellar evolution (e.g., Burrows \& Liebert
1993; Laughlin \& Bodenheimer 1993) and as a source of dark matter in
the galactic halo (e.g., Heygi \& Olive 1989; Adams \& Walker 1990;
Salpeter 1992; Graff \& Freese 1995).  In addition, microlensing
experiments are now providing an important probe of these low mass
stellar populations (e.g., Alcock et al. 1993; Aubourg et
al. 1993). Within the paradigm of star formation invoked here,
however, the formation of large numbers of brown dwarfs is difficult.
In the following discussion, we examine this statement in more detail
for both limits presented in \S 3 and \S 4.

In the limit that the effective sound speed is the most important
physical variable (\S 3), the SEMF implies that in order to form 
very low mass stars, the effective sound speed must be very small. 
However, even the lowest temperatures expected in present day 
molecular clouds $T \sim 10$ K lead to the formation of stars with
masses greater than the brown dwarf limit.  Roughly speaking, stars
need to become reasonably large in order to produce winds sufficiently
powerful to reverse the infall; objects with masses smaller than the
brown dwarf limit can only form within star forming regions with
very small mass infall rates (such a scenario has been advocated by
Lenzuni, Chernoff, \& Salpeter 1992; see also Zinnecker 1995).  
We thus expect brown dwarfs to be rare. 
 
In the opposite limit in which many different physical variables
conspire to produce a nearly log-normal distribution (\S 4),  the 
characteristic mass scale and total width must have given values  
($m_C \approx 0.1$ and $\sigbar \approx 1.6$) in order to be 
consistent with the observed IMF.  This result implies that the 
number of stars with masses less than $m_C$ (and hence objects 
below the brown dwarf limit) is very highly suppressed.  We 
stress that this claim is stronger than a blind extrapolation of
the observed IMF into the unknown: In the limit of large $n$, the
distribution approaches a log-normal form and, other than $m_C$ and
$\sigbar$, {\it there are no additional parameters to specify}.  
For the IMF of equation [4.14], the fraction of the total mass that
resides in stars with masses less than the brown dwarf limit (taken
here to be $m_{BD}$ = 0.08) is $\sim 5\%$.  Notice also that this
putative brown dwarf population corresponds to the low mass end 
of the usual stellar population and {\it not} the halo population.  
If brown dwarfs make up a substantial fraction of the mass of the 
galactic halo, then they must arise from a population with an IMF 
different from that of field stars. 

\bigskip 
\centerline{\it 6.5 Future Work} 
\medskip 

Many directions for future research along these lines remain.  
In this paper, we have presented a basic framework which can be 
used to calculate theoretical models of the IMF.  Thus far, we 
have considered only extremely simple models for both the SEMF
and the distribution of initial conditions.  Thus, essentially 
all steps of the calculation can be improved significantly.

For the SEMF, more elaborate models of the type considered here 
can be derived (see also Appendix A).  In addition, future work 
should use hydrodynamic simulations to study the manner in which 
protostellar outflows reverse the infall.  Such work will help 
determine or constrain the form of the SEMF. Finally, one can 
also work backwards from the observed distributions of initial 
conditions and see what type of SEMF is required to produce the 
observed IMF. 

For the distributions of initial conditions, both theoretical and 
observational approaches should be pursued.  Almost all of the 
relevant physical variables appearing in the SEMF can be measured 
in actual star forming regions.  The corresponding distributions 
can also be determined, e.g., the effective sound speed $a$
(see \S 2),  the rotation rate $\Omega$ (Goodman et al. 1993),
the wind efficiency parameters $\beta/\alpha\epsilon$ (Lada 1985), 
and the star/disk mass fraction $\gamma$ (Beckwith \& Sargent 1993; 
Beckwith et al. 1990; Adams, Emerson, \& Fuller 1990).  Although 
some data on these distributions currently exist, much more is 
necessary to fully understand the problem. In addition to finding 
more accurate descriptions of these distributions, future 
observational surveys can also determine how the distributions of 
the variables change from one star forming environment to another. 

The relative importance of the physical variables is given by the 
size of the contribution $\sigma_j^2$ to the total width of the
distribution (see Table 1).  The effective sound speed is thus 
the most important variable. The rotation rate $\Omega$, the wind
efficiency factors $\beta/\alpha\epsilon$, and star/disk mass 
fraction $\gamma$ are the next most important.  Future studies 
should prioritize their efforts accordingly, both for observational 
approaches (as described above), and theoretical studies, which 
we discuss next. 

The distributions of the physical variables can also be calculated
theoretically.  For example, the distribution of effective sound 
speed can be calculated from the combination of the line-width vs
density relationship (equation [3.1]) and the clump mass distribution
(equation [3.3]). These distributions, in turn, can be calculated from
MHD wave considerations (Fatuzzo \& Adams 1993; McKee \& Zweibel 1995)
and molecular cloud models (Norman et al. 1995). Similarly, theoretical 
models of the protostellar wind mechanism (Shu et al. 1994) and disk 
stability calculations (Laughlin 1994) will eventually determine the 
distributions of the dimensionless parameters $\alpha$, $\beta$,
$\gamma$, and $\epsilon$ appearing in the SEMF.  Calculations of 
this type remain in their infancy; much more work must be done 
in order to fully understand these distributions. 

In summary, we have presented a basic calculational framework 
which can be used to build theoretical models of the IMF. 
This approach is based on a SEMF and the underlying distributions 
of the physical variables which enter into the star formation 
problem.  The simplest cases of these models are presented here 
and show reasonable agreement with the observed IMF.  In the 
future, all steps of this calculation can be improved and we 
hope to eventually obtain a fundamental understanding of the IMF.

\vskip0.5truein
\centerline{Acknowledgements} 

We would especially like to thank Hans Zinnecker for many stimulating
discussions spanning many years and many continents.  We also thank
Gary Bernstein, Joel Bregman, Gus Evrard, Curtis Gehman, Paul Ho, Jim
Langer, Mario Mateo, Avery Meiksin, Joe Monaghan, Phil Myers, Steve
Myers, Colin Norman, Frank Shu, Joe Silk, Rick Watkins, and Simon
White for useful discussions.  Finally, we thank the referee -- John
Scalo -- for useful comments which improved the paper.  This work was
supported by an NSF Young Investigator Award, NASA Grant No. NAG
5-2869, and by funds from the Physics Department at the University of
Michigan.  Finally, we thank the Institute for Theoretical Physics at
U. C. Santa Barbara and NSF Grant No. PHY94-07194.

\newpage 
\centerline{\bf APPENDIX A: ALTERNATE MODEL FOR THE} 
\centerline{\bf SEMI-EMPIRICAL MASS FORMULA} 
\bigskip 

In this appendix, we discuss an alternate derivation of 
the semi-empirical mass formula of \S 2.  We show that 
the final result is very similar to that derived in the 
text.  We thus conclude that the general form of the SEMF
is fairly robust.

At a given time in the collapse, most of the newly falling 
material falls to radii near the centrifugal radius $R_C$. 
We therefore consider the condition that the ram pressure of the 
outflow is sufficiently strong to reverse the infall at the 
radius $R_C$.  This condition defines the effective end of 
the infall phase and can be written in the form 
$${\dot M_w} v_w = \delta {\dot M} v_C \, , \eqno({\rm A}1)$$
where $v_C$ $\sim (GM/R_C)^{1/2}$ is the infall speed at the 
centrifugal radius and where $\delta$ is a dimensionless 
parameter. 

We use this condition in place of equation [2.1] (the old defining
equation for the end of infall) and keep the remaining assumptions
concerning the scalings of outflow strengths, etc.  After some
rearrangement, we obtain the alternate scaling relation 
$$L_\ast \, M_\ast^{1/2} \, R_\ast^{1/2} \, = 
4 m_0 \gamma^{3/2} \delta \biggl( {\beta \over \alpha \epsilon} 
\biggr) {a^7 \over G^{3/2} \Omega} = \const \, {a^7 \over G^{3/2} 
\Omega}  \, , \eqno({\rm A}2)$$ 
where we have defined the parameter $\const \sim 500$. 
Once again, the stellar properties appear on the left hand 
side of the equation and the initial conditions appear on 
the right hand side. 

In dimensionless form, this transformation can be written 
$${\widetilde L} \, m^{1/2} \, {\widetilde R}^{1/2} \, 
= 2500 \const_3 \, a_{35}^7 \Omega_1^{-1} \, , \eqno({\rm A}3)$$ 
where we have introduced ${\widetilde R} \equiv R_\ast/(1 R_\odot)$ 
and $\const_3 \equiv \const/10^3$. 

Although this transformation appears to have a somewhat different 
form than that derived in \S 2 in the text, it leads to exactly 
the same scalings for the case of (low mass) young stellar objects 
with luminosity dominated by infall energy. In this case, the 
SEMF takes the form 
$$m = \con a_{35}^3 \Omega_1^{-2/3} \, , \eqno({\rm A}4)$$
where the parameter $\con$ contains all of the original 
parameters of the problem.  This form is exactly the same 
as that obtained in \S 2. 

\newpage 
\centerline{\bf APPENDIX B: CALCULATION OF THE}
\centerline{\bf EXPECTED WIDTH OF A LOG-NORMAL IMF}
\bigskip 

In this Appendix, we estimate the total width $\sigbar$ and the 
mass scale $m_C$ of the IMF based on the SEMF of \S 2 and 
observed distributions of the fundamental physical parameters. 

The most important physical variable is the effective sound speed, 
which has the largest contribution to the total width.  The distribution 
of the effective sound speed can be determined from the clump mass 
distribution [3.3] and the relation [3.2] between the effective 
sound speed and the clump mass. The result is 
$${d N \over d a} = {\cal N} a^{-[(p-1)q + 1]} 
\, , \eqno({\rm B}1)$$
where $\cal N$ is a normalization constant and $q \approx 4$ 
and $p \approx 3/2$ are the indices of the distributions 
[3.2] and [3.3].  Since the index appearing in the distribution [B1] 
is $\sim 3$, we need to introduce a lower cutoff $a_0$ to keep 
the distribution bounded.  We take $a_0 \approx$ 0.20 km/s, 
corresponding to the thermal sound speed at a temperature 
$T = 10$ K. Next, we let $x = \ln a/a_0$ and calculate the 
mean of the variable $x$: 
$$\xbar = (p-1)q \, \int_0^\infty \, x \, dx \, 
{\rm e}^{-(p-1)qx} \, = {1 \over (p-1)q} \, 
\, . \eqno({\rm B}2)$$
Thus, the relevant reduced variable $\xi_a$ which shows 
how the sound speed enters into the IMF is given by 
$$\xi_a = \mu (x - \xbar) = \mu \Bigl\{ x - 
{1 \over (p-1)q} \Bigr\} \, \, , \eqno({\rm B}3)$$ 
where $\mu$ is the exponent of the sound speed appearing 
in the SEMF ($\mu$ = 2 -- 3, depending on the mass range). 
Another straightforward integration then gives us the 
variance 
$$\sigma_a^2 = {\mu^2 \over (p-1)^2  q^2} \, 
\, \approx 1.56 \, , \eqno({\rm B}4)$$
where we have used $\mu$ = 2.5, $p$ = 1.5, and $q=4$ 
to obtain the numerical estimate. Thus, the sound speed
contributes a little over half of the total width of 
the distribution (recall that $\sigbar^2$ = 2.45). 

Similarly, we can calculate the appropriate mean value 
of the sound speed from this distribution. As described
in the text, we must calculate the average of the 
logarithm of the variable, and then exponentiate
the result.  We thus obtain
$$\abar = \exp [ \langle \ln a \rangle ] = a_0 
\exp [1/ (q (p-1) ) ] \approx 0.33 \, \, {\rm km} 
\, \, {\rm s}^{-1} \, . \eqno({\rm B}5)$$ 

Of the remaining variables, the dimensionless factors 
$\alpha$, $\beta$, $\gamma$, etc. appearing in the SEMF 
can all be treated the same approximate manner. Suppose 
a given variable $z_j$ varies by a factor ${\cal F}_j$ and 
enters into the SEMF with an exponent $\mu_j$.  Then, the 
quantity $\xi_j = \mu_j \ln z_j$ lies in the interval 
$[- \mu_j \ln {\cal F}_j, + \mu_j \ln {\cal F}_j ]$. 
The contribution to the total variance is then approximately 
given by 
$$\sigma_j^2 = \mu_j^2 \bigl[ \ln {\cal F}_j \bigr]^2 
\, . \eqno({\rm B}6)$$ 

In the following table, we list estimates for the exponents $\mu_j$
and the expected variance factors ${\cal F}_j$ for the variables in
the SEMF. The final columns show the contribution $\sigma_j^2$ to 
the total width of the distribution and the mean value (as defined 
by equation [4.5]) which determines the characteristic mass scale. 
For the exponents $\mu_j$, we use the the form [2.15a] for the SEMF. 
The distribution for the rotation rate $\Omega$ was derived from the
results of Goodman et al. (1993).  The distribution for the mechanical
luminosity factors ($\beta/\alpha \epsilon$) for protostellar outflows
was taken from Figure 7 of Lada (1985); here we consider the three 
wind parameters $\alpha$, $\beta$, and $\epsilon$ to be coupled (not 
independent) and hence described by a single distribution.  
The star/disk mass fraction $\gamma$ has been estimated from 
various calculations of the stability of self-gravitating star/disk 
systems (see, e.g., Adams, Ruden, \& Shu 1989; Shu et al. 1990; 
Laughlin 1994; Woodward, Tohline, \& Hashisu 1994). 
Finally, the efficiency factor $\eta$ for protostellar luminosities 
has been estimated from calculations of protostellar structure 
(e.g., Stahler, Shu, \& Taam 1980; Palla \& Stahler 1990, 1992) 
and from radiative transfer models of protostellar spectral energy 
distributions (Adams \& Shu 1986; Adams, Lada, \& Shu 1987; 
Kenyon, Calvet, \& Hartmann 1993). 

Using the values given in Table 1 and the SEMF, we can calculate the
total width and characteristic mass scale. We find that the calculated
width is $\sigbar \approx 1.81$, which is slightly higher than the
observed value of $\sigbar$ = 1.57.  Similarly, the calculated mass
scale is $m_C \approx$ 0.25, which is again higher than the observed
value of $m_C$ = 0.095.  Thus, the calculated values are approximately
correct, but still differ from the observed values by a significant
amount. In any case, this calculation is meant to be illustrative 
rather than definitive.  

\vskip0.5truein
\centerline{\bf Table 1. Mean and Variance of Fundamental Parameters} 
\nobreak 
$$\matrix{
{\rm parameter} & \mu_j & {\cal F}_j & \sigma_j^2 & {\bar \alpha} \cr 
   ------       &  ---  &   ----     &   ----     & ---- \cr
a_{35}          &  5/2  &   N.A.     &   1.56 & 0.94 \cr 
\Omega_1        &  2/3  &     3      &   0.54 & 1.5 \cr 
\beta/\alpha \epsilon & 1/3 & 10     &   0.60 & 100 \cr  
\gamma          &   1   &     2      &   0.48 & 0.5 \cr    
\delta          &  1/3  &     2      &   0.05 & 1.0 \cr    
\eta            &  1/3  &     2      &   0.05 & 0.5 \cr }$$

\newpage 
\centerline{\bf REFERENCES} 
\medskip 

\par\pp
Abt, H. A. 1983, {\sl A R A \& A}, {\bf 21}, 343 

\par\pp
Abt, H. A., \& Levy, S. G. 1976, {\sl ApJ S}, {\bf 30}, 273 

\par\pp
Adams, F. C. 1990, {\sl ApJ}, {\bf 363}, 578 

\par\pp
Adams, F. C. 1995, in Bottom of the Main Sequence -- and Beyond, 
ed. C. G. Tinney (Berlin: Springer-Verlag), p. 171 

\par\pp
Adams, F. C., Emerson, J. E., \& Fuller, G. A. 1990,
{\sl ApJ}, {\bf 357}, 606

\par\pp
Adams, F. C., Galli, D., Najita, J., Lizano, S., \& Shu, F. H. 1995,
{\sl ApJ}, in preparation

\par\pp 
Adams, F. C., Lada, C. J., \& Shu, F. H. 1987, {\sl ApJ}, {\bf 321}, 788 

\par\pp 
Adams, F. C., Ruden, S. P., \& Shu, F. H. 1989, {\sl ApJ}, {\bf 347}, 959 

\par\pp
Adams, F. C., \& Shu, F. H. 1986, {\sl ApJ}, {\bf 308}, 836  

\par\pp
Adams, F. C., \& Walker, T. P. 1990, {\sl ApJ}, {\bf 359}, 57 

\par\pp
Alcock, C. et al. 1993, {\sl Nature}, {\bf 365}, 621

\par\pp
Aubourg, E. et al. 1993, {\sl Nature}, {\bf 365}, 623

\par\pp 
Bally, J., \& Lada, C. J. 1983, {\sl ApJ}, {\bf 265}, 824 

\par\pp
Beckwith, S.V.W., Sargent, A. I., Chini, R. S., \& Gusten, R.
1990, {\sl AJ}, {\bf 99}, 924

\par\pp
Beckwith, S.V.W., \& Sargent, A. I. 1993, in Protostars and Planets III, 
ed. E. Levy \& J. Lunine (Tucson: University of Arizona Press), p. 521 

\par\pp
Blitz, L. 1993, in Protostars and Planets III, ed. E. Levy \& 
J. Lunine (Tucson: University of Arizona Press), p. 125 

\par\pp
Bodenheimer, P. 1978, {\sl ApJ}, {\bf 224}, 488 

\par\pp
Bodenheimer, P., Ruzmaikina, T., \& Mathieu, R. D. 1993, 
in Protostars and Planets III, ed. E. Levy \& J. Lunine 
(Tucson: University of Arizona Press), p. 367 

\par\pp
Bonnell, I.. \& Bastien, P. 1992, in {Complementary Approaches to 
Double \& Multiple Star Research; IAU Colloquium No. 135}, ed. 
H. McAlister (Provo: Pub. Astr. Soc. Pacific), p. 206 

\par\pp
Boss, A. P. 1992, in {Complementary Approaches to Double \& Multiple
Star Research; IAU Colloquium No. 135}, ed. H. McAlister
(Provo: Pub. Astr. Soc. Pacific), p. 195

\par\pp
Burrows, A., Hubbard, W. B., \& Lunine, J. I. 1989, {\sl ApJ}, 
{\bf 345}, 939 

\par\pp
Burrows, A., Hubbard, W. B., Saumon, D., \& Lunine, J. I. 
1993, {\sl ApJ}, {\bf 406}, 158 

\par\pp
Burrows, A., \& Liebert, J. 1993, {\sl Rev. Mod. Phys.}, {\bf 65}, 301

\par\pp
Cassen P., \& Moosman, A. 1981, {\sl Icarus}, {\bf 48}, 353 

\par\pp
Clarke, C. J., \& Pringle, J. E. 1991, {\sl M N R A S}, {\bf 249}, 588

\par\pp
Dickman, R. L., Horvath, M. A., \& Margulis, M. 1990,
{\sl ApJ}, {\bf 365}, 586

\par\pp
Duerr, R., Imhoff, C. L., \& Lada, C. J. 1982, {\sl ApJ}, {\bf 261}, 135 

\par\pp
Duquennoy, A., \& Mayor, M. 1991, {\sl A \& A}, {\bf 248}, 485

\par\pp
Edwards, S., Ray, T., \& Mundt, R. 1993, in Protostars and Planets III,
ed. E. Levy \& J. Lunine (Tucson: University of Arizona Press), p. 567 

\par\pp
Elmegreen, B. G., \& Mathieu, R. D. 1983, {\sl MNRAS}, {\bf 203}, 305 

\par\pp
Elmegreen, B. G. 1985, in Birth and Infancy of Stars, eds. 
R. Lucas, A. Omont, \& R. Stora (Amsterdam: North Holland), p. 257 

\par\pp
Fatuzzo, M., \& Adams, F. C. 1993, {\sl ApJ}, {\bf 412}, 146 

\par\pp
Fletcher, A. B., \& Stahler, S. W. 1994, {\sl ApJ}, {\bf 435}, 313 

\par\pp
Goodman, A. A., Benson, P. J., Fuller, G. A., \& Myers, P. C. 
1993, {\sl ApJ}, {\bf 406}, 528 

\par\pp
Graff, D., \& Freese, K. 1995, {\sl ApJ Letters}, in press 

\par\pp
Heygi, D. J., \& Olive, K. A. 1989, {\sl ApJ}, {\bf 346}, 648 

\par\pp
Hillenbrand, L. A., Massey, P., Strom, S. E., \& Merrill, K. M. 
1993, {\sl AJ}, {\bf 106}, 1906 

\par\pp
Houlahan, P., \& Scalo, J. M. 1992, {\sl ApJ}, {\bf 393}, 172 

\par\pp
Hoyle, F. 1953, {\sl ApJ}, {\bf 118}, 513 

\par\pp
Jijina, J., \& Adams, F. C. 1995, {\sl ApJ}, in press 

\par\pp
Kenyon, S. J., Calvet, N., \& Hartmann, L. 1993,
{\sl ApJ}, {\bf 414}, 676

\par\pp
K{\" o}nigl, A., \& Ruden, S. P. 1993, in Protostars and Planets III,
ed. E. Levy \& J. Lunine (Tucson: University of Arizona Press), p. 641  

\par\pp 
Lada, C. J. 1985, {\sl A R A \& A}, {\bf 23}, 267 

\par\pp 
Lada, C. J., \& Shu, F. H. 1990, {\sl Science}, {\bf 1111}, 1222

\par\pp
Lada, E. A., Bally, J., \& Stark, A. A. 1991, {\sl ApJ}, {\bf 368}, 444

\par\pp
Lada, E. A., Strom, K. M., \& Myers, P. C. 1993, 
in Protostars and Planets III, ed. E. Levy \& J. Lunine 
(Tucson: University of Arizona Press), p. 245

\par\pp
Larson, R. B. 1973, {\sl MNRAS}, {\bf 161}, 133 

\par\pp
Larson, R. B. 1981, {\sl MNRAS}, {\bf 194}, 809 

\par\pp
Larson, R. B. 1992, {\sl MNRAS}, {\bf 256}, 641 

\par\pp
Larson, R. B. 1995, {\sl MNRAS}, {\bf 272}, 213 

\par\pp
Laughlin, G. P. 1994, {\sl PhD Thesis: The Formation and Evolution
of Protostellar Disks}, University of California, Santa Cruz

\par\pp
Laughlin, G., \& Bodenheimer, P. 1993, {\sl ApJ}, {\bf 403}, 303 

\par\pp
Lenzuni, P., Chernoff, D. F., \& Salpeter, E. E. 1992, 
{\sl ApJ}, {\bf 393}, 232

\par\pp
Levreault, R. M. 1988, {\sl ApJ}, {\bf 330}, 897 
     
\par\pp 
Lizano, S., \& Shu, F. H. 1989, {\sl ApJ}, {\bf 342}, 834 

\par\pp
Massey, P., Lang, C. C., DeGioia-Eastwood, K., \& Garmany, C. D. 
1995, {\sl ApJ}, {\bf 438}, 188  

\par\pp
McKee, C. F., \& Zweibel, E. G. 1995, {\sl ApJ}, {\bf 440}, 686

\par\pp
Miller, G. E., \& Scalo, J. M. 1979, {\sl ApJ Suppl.}, {\bf 41}, 513

\par\pp
Myers, P. C., \& Fuller, G. A. 1992, {\sl ApJ}, {\bf 396}, 631 

\par\pp
Myers, P. C., \& Fuller, G. A. 1993, {\sl ApJ}, {\bf 402}, 635 

\par\pp
Nakano, T. 1989, {\sl ApJ}, {\bf 345}, 464 

\par\pp
Nakano, T., Hasegawa, T., \& Norman, C. 1995, {\sl ApJ}, {\bf 450}, 183 

\par\pp
Norman, C. A., Adams, F. C., Cowley, S., \& Sudan, R. 1995, 
{\sl ApJ}, to be submitted 

\par\pp
Palla, F., \& Stahler, S. W. 1990, {\sl ApJ}, {\bf 360}, L47 

\par\pp
Palla, F., \& Stahler, S. W. 1992, {\sl ApJ}, {\bf 392}, 667

\par\pp
Parzen, E. 1960, Modern Probability Theory and Its Applications
(New York: Wiley) 

\par\pp
Phillips, A. C. 1994, The Physics of Stars (Chichester: Wiley) 

\par\pp
Rana, N. C. 1991, {\sl A R A \& A}, {\bf 29}, 129 

\par\pp
Richtmyer, R. D. 1978, {Principles of Advanced Mathematical Physics} 
(New York: Springer-Verlag) 


\par\pp
Salpeter, E. E. 1955, {\sl ApJ}, {\bf 121}, 161 

\par\pp
Salpeter, E. E. 1992, {\sl ApJ}, {\bf 393}, 258

\par\pp
Scalo, J. M. 1985, in Protostars and Planets II, ed. D. C. 
Black \& M. S. Mathews (Tucson: Univ. Arizona Press), 201 

\par\pp
Scalo, J. M. 1986, {\sl Fund. Cos. Phys.}, {\bf 11}, 1 

\par\pp
Scalo, J. M. 1987, in {\sl Interstellar Processes}, ed. 
D. J. Hollenbach \& H. A. Thronson (Dordrecht: Reidel), p. 349

\par\pp
Scalo, J. M. 1990, in {Physical Processes in Fragmentation
and Star Formation}, eds. R. Capuzzo-Dolcetta et al.
(Dordrecht: Kluwer), p. 151

\par\pp 
Shu, F. H. 1977, {\sl ApJ}, {\bf 214}, 488 

\par\pp 
Shu, F. H., Adams, F. C., \& Lizano, S. 1987, 
{\sl A R A \& A}, {\bf 25}, 23 (SAL) 

\par\pp 
Shu, F. H., Lizano, S., \& Adams, F. C. 1987, in Star Forming 
Regions, IAU Symp. No. 115, ed. M. Peimbert \& J. Jugaku 
(Dordrecht: Reidel), p. 417 (SLA) 

\par\pp
Shu, F. H., Lizano, S., Ruden, S. P., \& Najita, J. 1988, 
{\sl ApJ}, {\bf 328}, L19 

\par\pp
Shu, F. H., Lizano, S., Ruden, S. P., \& Najita, J. 1994,
{\sl ApJ}, {\bf 429}, 797
     
\par\pp 
Shu, F. H., Tremaine, S., Adams, F. C., \& Ruden, S. P. 1990, 
{\sl ApJ}, {\bf 358}, 495 

\par\pp 
Silk, J. 1995, {\sl ApJ}, {\bf 438}, L41 
     
\par\pp 
Stahler, S. W. 1983, {\sl ApJ}, {\bf 274}, 822 

\par\pp 
Stahler, S. W. 1988, {\sl ApJ}, {\bf 332}, 804 

\par\pp 
Stahler, S. W., Shu, F. H., \& Taam, R. E. 1980, {\sl ApJ}, 
{\bf 241}, 63 

\par\pp
Tatematsu, K. et al. 1993, {\sl ApJ}, {\bf 404}, 643 

\par\pp 
Terebey, S., Shu, F. H., \& Cassen, P. 1984, {\sl ApJ}, {\bf 286}, 529 

\par\pp
Tinney, C. G. 1995, editor, Bottom of the Main Sequence -- and Beyond
(Berlin: Springer-Verlag) 

\par\pp
Williams, J. P., de Geus, E. J., \& Blitz, L. 1994, 
{\sl ApJ}, {\bf 428}, 693 

\par\pp
Wolfire, M. G., \& Cassinelli, J. P. 1986, {\sl ApJ}, {\bf 310}, 207 

\par\pp
Wolfire, M. G., \& Cassinelli, J. P. 1987, {\sl ApJ}, {\bf 319}, 850 

\par\pp
Woodward, J. W., Tohline, J. E., \& Hashisu, I. 1994,
{\sl ApJ}, {\bf 420}, 247 

\par\pp
Zinnecker, H. 1984, {\sl M N R A S}, {\bf 210}, 43 

\par\pp
Zinnecker, H. 1985, in Birth and Infancy of Stars, eds. 
R. Lucas, A. Omont, \& R. Stora (Amsterdam: North Holland), p. 473 

\par\pp
Zinnecker, H. 1989, in Evolutionary Phenomena in Galaxies, eds. 
J. E. Beckman \& B.E.J. Pagel (Cambridge: Cambridge University 
Press), p. 113 

\par\pp
Zinnecker, H. 1990, in Physical Processes in Fragmentation and 
Star Formation, eds. R. Capuzzo-Dolcetta, C. Chiosi, \& A. Di Fazio
(Dordrecht: Kluwer), p. 201 

\par\pp
Zinnecker, H. 1995, in Bottom of the Main Sequence -- and Beyond, 
ed. C. G. Tinney (Berlin: Springer-Verlag), p. 257

\par\pp
Zinnecker, H., McCaughrean, M. J., \& Wilking, B. A. 1993, 
in Protostars \& Planets III, ed. E. Levy \& J. Lunine 
(Tucson: University of Arizona Press), p. 429 

\par\pp
Zuckerman, B., \& Palmer, P. 1974, {\sl A R A \& A}, {\bf 12}, 279 

\vskip 1.75truein
\bigskip 
\centerline{\bf FIGURE CAPTIONS} 
\medskip 

\medskip 
Figure 1. Observed estimates of the initial mass function. 
Dashed curve shows the power-law IMF of Salpeter (1955). 
Solid curve shows the log-normal analytic fit to the 
IMF from Miller \& Scalo (1979).  The dashed curve with 
symbols shows a more recent empirical estimate of the 
IMF taken from Rana (1991).  All three distributions are
normalized to unity at $m=1$ ($M_\ast = 1 M_\odot$).  

\medskip 
Figure 2. The masses of forming stars as a function of initial 
conditions for the SEMF of \S 2. This figure shows contours of 
constant mass in the plane of initial conditions.  The effective 
sound speed $a_{35}$ = $a$/(0.35 km s$^{-1}$) constitutes the 
horizontal axis and the rotation rate 
$\Omega_1 = \Omega/$(1 km s$^{-1}$ pc$^{-1}$)  
constitutes the vertical axis.  The remaining parameters of the 
problem are taken to have constant values as described in the text. 
The contours correspond to masses in the range $0.1 \le m \le 100$, 
with the mass increasing from left to right in the figure. 
The region in the upper left corner (above the dashed curve) 
corresponds to brown dwarfs; the region in the lower right corner 
corresponds to stars so massive that they become unstable. 

\medskip 
Figure 3. Empirical model for the initial mass function.
Dashed curve shows the IMF resulting from the semi-empirical 
mass formula of \S 2 and the observed scaling laws which 
describe the distribution of effective sound speed, i.e., 
the distribution of initial conditions. 
The solid curve shows the Miller/Scalo fit to the observed IMF.  
Both curves are normalized to unity at $m=1$ ($M_\ast = 1 M_\odot$).  

\medskip 
Figure 4. Random model for the initial mass function.
Dashed curve shows the IMF resulting from the semi-empirical 
mass formula of \S 2 and a distribution of initial conditions 
described by a collection of $n=10$ random variables. 
The solid curve shows the Miller/Scalo fit to the observed IMF.  
Both curves are normalized to unity at $m=1$ ($M_\ast = 1 M_\odot$).  

\medskip 
Figure 5. Random model for the IMF for different numbers $n$ of 
the fundamental variables.  The solid curve shows the Miller/Scalo 
fit to the observed IMF.  Dashed curves show distributions 
calculated from a collection of random variables with $n$ = 
1, 2, 3, and 5. As the value of $n$ increases, the curves become 
closer to the log-normal ($n \to \infty$) limit.  All curves have 
been scaled so that the total width $\sigbar$ of the distribution 
and the characteristic mass scale $m_C$ agree with the observed 
values from the Miller/Scalo IMF, i.e., $m_C = 0.095$ and
$\sigbar = 1.57$.  In addition, all curves are normalized 
to unity at $m=1$ ($M_\ast = 1 M_\odot$). 

\medskip 
Figure 6. Composite distributions for the IMF using $n$ 
fundamental variables with power-law distributions. 
The various curves are shown for $n$ = 1, 3, 10 and 20. 
Also shown is the log-normal curve corresponding to the 
limit $n \to \infty$.  All curves have been scaled so that 
the total width $\sigbar$ of the distribution and the 
characteristic mass scale $m_C$ agree with the observed 
values from the Miller/Scalo IMF, i.e., $m_C = 0.095$ and 
$\sigbar = 1.57$.  In addition, all curves are normalized 
to unity at $m=1$ ($M_\ast = 1 M_\odot$).  

\medskip 
Figure 7. The effects of binary companions on the IMF. Solid curve 
shows the primary mass distribution (taken here to have the 
Miller/Scalo form). Dashed curve shows the effects of including an 
additional distribution of binary companions, where we have used the 
prescription of \S 5.3. The binary fraction $\cal F$ = 0.75, the 
variance of the mass ratio distribution is $\sigma_{\alpha S}$ = 1, 
and the mean of the mass ratio distribution is 
$\langle \ln \alpha_S \rangle$ = $-1$.  Both curves 
are normalized to unity at $m=1$ ($M_\ast = 1 M_\odot$).  

\bye